\documentclass{aa} 
\usepackage[varg]{txfonts}
\usepackage{graphicx}

\usepackage[colorlinks,citecolor=blue,bookmarks=false,breaklinks=true]{hyperref}

\usepackage{natbib}
\bibpunct{(}{)}{;}{a}{}{,} 

\usepackage{multirow}
\usepackage{makecell}

\begin{document}

\title{Compact dust-obscured star formation \\ and the origin of the galaxy bimodality}

\author{Maxime Tarrasse\inst{\ref{inst1}}
   \and Carlos Gómez-Guijarro\inst{\ref{inst1}}
   \and David Elbaz\inst{\ref{inst1}}
   \and Benjamin Magnelli\inst{\ref{inst1}}
   \and Mark Dickinson\inst{\ref{inst2}}
   \and {Aur{\'e}lien Henry}\inst{\ref{inst3}}
   \and Maximilien Franco\inst{\ref{inst1},\ref{inst4}}
   \and Yipeng Lyu\inst{\ref{inst1}}
   \and Jean-Baptiste Billand\inst{\ref{inst1}}
   \and Rachana Bhatawdekar\inst{\ref{inst5}}
   \and Yingjie Cheng\inst{\ref{inst6}}
   \and Adriano Fontana\inst{\ref{inst7}}
   \and Steven L. Finkelstein\inst{\ref{inst4}}
   \and Giovanni Gandolfi\inst{\ref{inst8},\ref{inst9}}
   \and Nimish Hathi\inst{\ref{inst10}}
   \and Michaela Hirschmann\inst{\ref{inst11},\ref{inst12}}
   \and Benne W. Holwerda\inst{\ref{inst13}}
   \and Anton M. Koekemoer\inst{\ref{inst10}}
   \and Ray A. Lucas\inst{\ref{inst10}}
   \and {Lise-Marie Seillé}\inst{\ref{inst14}}
   \and Stephen Wilkins\inst{\ref{inst15},\ref{inst16}}
   \and  {{L. Y. Aaron} {Yung}}\inst{\ref{inst10}}
   }

   \institute{Université Paris-Saclay, Université Paris Cité, CEA, CNRS, AIM, 91191, Gif-sur-Yvette, France\label{inst1}
   \and NSF’s National Optical-Infrared Astronomy Research Laboratory, 950 N. Cherry Ave., Tucson, AZ 85719, USA\label{inst2}
   \and Department of Physics, University of California, Merced, 5200 Lake Road, Merced, CA 92543, USA\label{inst3}
   \and The University of Texas at Austin, 2515 Speedway Blvd Stop C1400, Austin, TX 78712, USA\label{inst4}
   \and European Space Agency (ESA), European Space Astronomy Centre (ESAC), Camino Bajo del Castillo s/n, 28692 Villanueva de la Cañada, Madrid, Spain\label{inst5}
   \and University of Massachusetts Amherst, 710 North Pleasant Street, Amherst, MA 01003-9305, USA\label{inst6}
   \and INAF Osservatorio Astronomico di Roma, Via Frascati 33, 00078 Monteporzio Catone, Rome, Italy\label{inst7}
   \and Dipartimento di Fisica e Astronomia "G.Galilei", Universit\'a di Padova, Via Marzolo 8, I-35131 Padova, Italy\label{inst8}
   \and INAF--Osservatorio Astronomico di Padova, Vicolo dell'Osservatorio 5, I-35122, Padova, Italy\label{inst9}
   \and Space Telescope Science Institute (STSCI), 3700 San Martin Dr., Baltimore, 21218 MD, USA\label{inst10}
   \and {\'E}cole Polytechnique Fed{\'e}rale de Lausanne (EPFL), Observatoire de Sauverny, Chemin Pegasi 51, CH-1290 Versoix, Switzerland\label{inst11}
   \and INAF Astronomical Observatory of Trieste, via G.B. Tiepolo 11, I-34143 Trieste, Italy\label{inst12}
   \and Department of Physics and Astronomy, University of Louisville, Natural Science Building 102, 40292 KY, Louisville, USA\label{inst13}
   \and Aix Marseille Univ, CNRS, CNES, LAM, Marseille, France\label{inst14}
   \and Astronomy Centre, University of Sussex, Falmer, Brighton BN1 9QH, UK\label{inst15}
   \and Institute of Space Sciences and Astronomy, University of Malta, Msida, MSD 2080, Malta\label{inst16}
}

\date{Received 4 November 2024 / Accepted 23 March 2025}


\abstract 
{The combined capabilities of the \textit{James Webb} Space Telescope/Near Infrared Camera (NIRCam) and the \textit{Hubble} Space Telescope/Advanced Camera for Surveys (ACS) instruments provide high-angular-resolution imaging from the ultraviolet to near-infrared (UV/NIR), offering unprecedented insight into the inner structure of star-forming galaxies (SFGs) even when they are shrouded in dust. In particular, it is now possible to spatially resolve and study a population of highly attenuated and massive red SFGs ($\rm RedSFGs$) at $z\sim4$ in  the rest-frame optical/near-infrared (optical/NIR). Given their significant contribution to the cosmic star formation rate density (SFRD) at $z>3$, these $\rm RedSFGs$ are likely to be the progenitors of the massive (${\rm log}(M_\ast/M_\odot) > 10$) and passive galaxies already in place at cosmic noon ($z\sim2)$. They therefore represent a crucial population that can help elucidate the mechanisms governing the transition from vigorous star formation to quiescence at high redshifts.} 
{We assembled a mass-complete sample of massive galaxies at $z=3-4$  to study and compare the stellar mass, star formation rate (SFR), dust attenuation, and age spatial distributions of $\rm RedSFGs$ with those of quiescent galaxies (QGs) and more typical blue SFGs ($\rm BlueSFGs$).} 
{We performed an injection-recovery procedure with galaxies of various profiles in the CEERS images to build a mass-complete sample of 188 galaxies with ${\rm log}(M_\ast/M_\odot) > 9.6,$ which we classified into $\rm BlueSFGs$, $\rm RedSFGs,$ and QGs. We performed a resolved spectral energy distribution (SED) fitting on the UV/NIR data to compute and compare the radial profiles of these three populations.} 
{The $\rm RedSFGs$ fraction is systematically higher than that of QGs and both are seen to increase with stellar mass. Together, they account for more than $50\%$ of galaxies with ${\rm log}(M_\ast/M_\odot)> 10.4$ at this redshift. This transition mass corresponds to the ${\rm log}(M_\ast/M_\odot)\sim10.4$ threshold, often referred to as the "critical mass," which delineates the bimodality between $\rm BlueSFGs$ and QGs. We find that $\rm RedSFGs$ and QGs present similar stellar surface density profiles and that $\rm RedSFGs$ manifest a dust attenuation concentration that is significantly higher than that of $\rm BlueSFGs$ at all masses. This suggests that a path for a $\rm BlueSFG$ to become quiescent is through a major compaction event, triggered once the galaxy reaches a sufficient mass, leading to the in situ formation of a massive bulge.}
{There is a bimodality between extended $\rm BlueSFGs$ and compact and strongly attenuated $\rm RedSFGs$ that have undergone a phase of major gas compaction. There is evidence that this early-stage separation is at the origin of the local bimodality between $\rm BlueSFGs$ and QGs, which we refer to as a "primeval bimodality."}

\titlerunning{Compact dust-obscured star formation and the origins of  galaxy bimodality}
\authorrunning{Tarrasse et al.}

\maketitle
\nolinenumbers

\section{Introduction}
Over the past few decades, our understanding of galaxy evolution has been greatly enhanced by multiwavelength observations. In particular, the identification of a tight ($<0.3$ dex) relation between the stellar masses and the star formation rates (SFRs) of star-forming galaxies \citep[the so-called star-formation main sequence, MS;][]{daddiMultiwavelengthStudyMassive2007,elbazReversalStarFormationdensity2007,noeskeStarFormationAEGIS2007,schreiberHerschelViewDominant2015} at least up to $z=6$ \citep{popessoMainSequenceStarforming2023,calabroEvolutionStarFormation2024}, with hints of its presence up to $z=9$ \citep{cieslaIdentificationTransitionStochastic2024}. This scenario suggests that the bulk of present-day stars formed through a secular process rather than through a series of bursts. Nevertheless, quiescent galaxies (QGs) do not follow this correlation and are located below the MS, indicating a suppressed star-formation for this population. This change in the specific star formation (${\rm sSFR\equiv SFR}/M_\ast$) is not fully understood and corresponds to the clear color bimodality observed between QGs and SFGs, with the latter having typically blue continuum-dominated spectra produced by massive young stars, while the former have redder spectra, dominated by the emission of older, less massive stars \citep{baldryQuantifyingBimodalColorMagnitude2004,williamsDetectionQuiescentGalaxies2009,patelUVJSelectionQuiescent2012}. This color bimodality also seems to be accompanied by a morphological bimodality linked to their morphological evolution \citep{leeCANDELSCORRELATIONGALAXY2013,osborneCANDELSMeetsGSWLC2020,liuMorphologicalTransformationQuenching2019}, with SFGs exhibiting disky structures (Sérsic index, $n\sim 1$); whereas QGs have more compact morphologies, with a Sérsic index of $n\sim 4$ \citep{vanderwel3DHSTCANDELSEvolution2014,wardEvolutionSizeMass2024}. This morphological and color bimodality has been shown to arise at high masses \citep{wuytsGalaxyStructureMode2011,huertas-companyGalaxyMorphologyLens2024}, suggesting that the rise in this bimodality is more of a mass-driven process, rather than an SFR- or morphology-driven one \citep{conseliceEPOCHSPaperDependence2024}. Additionally, it was shown that the high-mass end ${\rm log}(M_\ast/M_\odot)>10$ of the cosmic stellar mass density (SMD) is dominated by QGs for redshifts lower than $z=2$ \citep{brammerNumberDensityMass2011,muzzinEVOLUTIONSTELLARMASS2013,weaverCOSMOS2020GalaxyStellar2023}. As this redshift defines the peak of the SFRD of the Universe \citep[also known as cosmic noon;][]{madauCosmicStarFormationHistory2014,matthewsCosmicStarFormation2021,leroyRoleStellarMass2024} and coincides with a significant fraction of QGs in place at this epoch \citep{brammerNumberDensityMass2011,vanderwel3DHSTCANDELSEvolution2014,vandokkumFormingCompactMassive2015,carnallSurprisingAbundanceMassive2023}, it implies that this bimodality between SFGs and QGs is forged early on in its history. Therefore, studying the possible progenitors of massive QGs already in place at $z\sim 2$ is essential to our understanding of how SFGs transition to quiescence.

In this context, the combined capabilities of the \textit{Spitzer} Space Telescope (\textit{Spitzer}) and the Atacama Large Millimeter/submillimeter Array (ALMA), has shed light on a population of galaxies extremely faint from the rest-frame ultraviolet (UV) up to the optical, making them undetectable, even in the reddest and deepest images of the \textit{Hubble} Space Telescope \citep[HST;][]{caputiNatureExtremelyRed2012,wangInfraredColorSelection2016,francoGOODSALMAMmGalaxy2018,wangDominantPopulationOptically2019,yamaguchiALMA26Arcmin22019,williamsDiscoveryDarkMassive2019}. This population was seen as a serious candidate for acting as the progenitor of the massive QG population observed at high redshifts \citep{carnallSurprisingAbundanceMassive2023}, given their significant contribution to the high-mass end (${\rm log}(M_\ast/M_\odot)>10.3$) of the cosmic SFRD and SMD at $3<z<6$, and their typical halo mass (${\sim} 10.13~h^{-1} M_\odot$ at $z=4$), which aligns with this hypothesis \citep{wangDominantPopulationOptically2019,gruppioniALPINEALMACIISurvey2020}. To detect this population, \citet{wangDominantPopulationOptically2019} used the $H$-dropout selection technique, which consists of searching for sources with very weak detection in the $H$-band (typically $H > 27$ mag) and bright detection in the \textit{Spitzer}/IRAC [$4.5~\mu$m]-band (typically $[4.5]<24$). Because of these observed properties, this selection resulted in a population of $z>3$ massive SFGs, heavily obscured by dust. These galaxies are referred  to as "HST-dark," "$H$-dropouts," or "optically dark" galaxies.

To fill the gap between Lyman break galaxies \citep[LBGs;][]{giavaliscoRestFrameUltravioletLuminosity2004,debarrosProperties36Lyman2014, arrabalharoDifferencesSimilaritiesStellar2020} and $H$-dropouts at $3<z<6$, \citet{xiaoHiddenSideCosmic2023} extended the HST-dark population to a wider selection ($H > 26.5$ and $[4.5]<24$). This provided a more complete view of SFRD at these redshifts by considering galaxies that are less extreme in terms of dust attenuation and stellar mass. The first studies of this so-called optically-faint galaxy (OFG) population using photometric and ALMA data have located it at the high-mass end of the MS (${\rm log}(M_\ast/M_\odot)>10$). It is characterized by a particularly high dust attenuation ($A_{\rm v}>2$), high star formation efficiency (SFE), and a low gas fraction, suggesting a highly efficient conversion of cold gas into stars at work in these systems \citep{alcaldepampliegaOpticallyFaintMassive2019, gomez-guijarroGOODSALMASourceCatalog2022, xiaoHiddenSideCosmic2023, gomez-guijarroJWSTCEERSProbes2023,xiaoMassiveOpticallyDark2023}. Nevertheless, \textit{Spitzer}'s low angular resolution in the NIR have not permitted a thorough morphological study of this population or of galaxies in general at these redshifts.

With the launch of the \textit{James Webb} Space Telescope (JWST), astronomers now have access to  unprecedentedly deep and high angular-resolution near-infrared (NIR) imaging at wavelengths longer than 2 $\rm \mu m$. This allows  us to spatially resolve the stellar structures of galaxies across cosmic time \citep{gardnerJamesWebbSpace2006,gardnerJamesWebbSpace2023}. While the definition of this population of red, dust-obscured star-forming galaxies, heavily depend on the color criteria used, the first studies using JWST NIR observations have confirmed their high contribution to the high-mass end of the SFRD (${\sim} 3.2 \times 10^{-3} M_\odot~\rm{yr^{-1} Mpc^{-3}}$) from $z=3$ up to $z=7$ \citep{barrufetUnveilingNatureInfrared2023} and total SMD, with the latter having been found to be underestimated in the pre-JWST area by ${\sim} 20\%$ for $3<z<4$ and up to ${\sim} 40\%$ for $6<z<8$ \citep{gottumukkalaUnveilingHiddenUniverse2024,weibelGalaxyBuildfirst152024}. Additionally, spectroscopic and optical-to-mid-IR (optical/MIR) observations of some of these sources have confirmed their high redshift, massive and dusty properties, while discussing the potential triality of their nature; namely, star-forming (${\sim} 70\%$), quiescent (${\sim} 20\%$), potential active galactic nuclei (AGNs) or young starbursts (${\sim} 10\%$) at $2<z<7$ \citep{perez-gonzalezCEERSKeyPaper2023,barrufetQuiescentDustyUnveiling2024}. A few papers have used the resolving power in the NIR to study the internal structures of these galaxies, highlighting their stellar and dust concentrated profiles \citep{perez-gonzalezCEERSKeyPaper2023,nelsonJWSTRevealsPopulation2023,gomez-guijarroJWSTCEERSProbes2023,smailHiddenGiantsJWST2023, sunJADESResolvingStellar2024}. Although these new studies seem to link this population even more strongly to the QGs, their spatially resolved properties are still poorly understood as they only focus on small samples and suffer from completeness bias. We therefore lack morphological evidence that this population corresponds to a key transition phase between SFGs and QGs.

In this work, we study the spatial distribution of different physical quantities for a mass-complete sample at $3<z<4$, using a resolved SED-fitting procedure. We computed and compared the radial profiles of color-selected QGs, typical blue SFGs (hereafter, referred to as $\rm BlueSFGs$) and red SFGs (hereafter, $\rm RedSFGs$) which include, but are not limited to, OFGs (see selection in Fig.~\ref{cleaned_OFG_selection.pdf}). In this way we aim to decipher the nature and potential role of the $\rm RedSFG$ population in the morphological transition from $\rm BlueSFGs$ to QGs.

This paper is organized as follows: In Sect.~\ref{Observations and Data Description}, we introduce our data, which includes JWST/NIRCam \citep{riekeNGSTNIRCamScientific2003,riekeOverviewJamesWebb2005,beichmanScienceOpportunitiesIR2012,riekePerformanceNIRCamJWST2023} and HST/ACS \citep{fordAdvancedCameraHubble1998,ryonACSInstrumentHandbook2023} observations from the CEERS field, and the associated catalog. We also present our SED fitting analysis, which we use for both resolved and unresolved studies. We take particular care to explain the choice of attenuation law for this work. Section~\ref{Determination and classification of the galaxies of the sample} is dedicated to the method used to assemble a mass-complete sample of $\rm BlueSFGs$, $\rm RedSFGs$ and QGs. In Sect.~\ref{Results}, we analyze the results of this resolved SED fitting method and, in particular, we compare the radial profiles of the physical properties of $\rm BlueSFGs$, $\rm RedSFGs$, and QGs. We discuss in Sect.~\ref{Discussion} how these results highlight the role of the $\rm RedSFG$ population in galaxy evolution by integrating our results into the broader framework of galaxy evolution. Finally, Sect.~\ref{Summary} summarizes our findings.
Throughout this work, we adopt a flat cosmology with [$\Omega_{\Lambda}, \Omega_{\rm M}] = [0.7, 0.3]$ and set the Hubble constant to $H_{0}=70~{\rm km.s^{-1}}$. We also quote all magnitudes in the AB system \citep{okeSecondaryStandardStars1983}.

\section{Observations and data description}
\label{Observations and Data Description}

\subsection{CEERS data}

In this study, we used imaging data obtained by the JWST/NIRCam instrument as part of the Cosmic Evolution Early Released Science survey \citep[CEERS;][]{finkelsteinCosmicEvolutionEarly2017}. It has been conducted in the Extended Groth Strip field (EGS) and contains ten pointings, covering a total area of 97 square arcminutes. CEERS provided the first deep extragalactic survey with JWST in seven NIRCam filters ($F115W$, $F150W$, $F200W$, $F277W$, $F356W$, $F410M$, and $F444W$) with a median $5\sigma$ point source depth of 29.20, 29.01, 29.16, 29.15, 29.15, 28.32, and 28.56, respectively. Additionally, we used the $F606W$ and $F814W$ optical-bands from HST observations with the ACS/WFC instrument, having a $5\sigma$ point source depth of 28.73 and 28.50, respectively \citep{davisAllWavelengthExtendedGroth2007,finkelsteinCompleteCEERSEarly2024}.

The JWST/NIRCam data have been reduced following \citet{bagleyCEERSEpochNIRCam2023} and the resulting images were pixel aligned to the HST/ACS images of the Cosmic Assembly Near-infrared Deep Extragalactic Legacy Survey \citep[CANDELS;][]{koekemoerCANDELSCosmicAssembly2011a,groginCANDELSCOSMICASSEMBLY2011}. In brief, pointings from June 2022 were processed using the JWST Calibration pipeline v1.7.2 and CRDS pmap 0989, while those from December 2022 were processed using the JWST Calibration pipeline v1.8.5 and CRDS pmap 1023. Then, a customized pipeline developed by the CEERS team incorporating procedures such as $1/f$ noise subtraction and artifact removal was applied to all pointings. Finally, the mosaics were aligned using astrometry data from Gaia-EDR3 \citep{gaiacollaborationGaiaEarlyData2021}, with a pixel scale of 0.03 arcsec/pixel, and background subtracted.

\subsection{Catalog}
\label{Catalog}

We used the photometric catalog built by \citet{gomez-guijarroJWSTCEERSProbes2023}, which contains our nine HST/JWST bands of interest ($F606W$, $F814W$, $F115W$, $F150W$, $F200W$, $F277W$, $F356W$, $F410M$, and $F444W$) together with the Canada-France-Hawaii Telescope (CFHT)/MegaCam \citep{bouladeMegacamNextgenerationWidefield1998} observations imaging in the EGS field in the $\rm u^\star, g^\prime, r^\prime, i^\prime$, and $\rm z^\prime$-bands \citep{gwynCanadaFranceHawaiiTelescopeLegacy2012}. This cataloging analysis provided segmentation maps, fluxes, and flux uncertainties for each source across the ten CEERS pointings in both the two HST and seven JWST bands used in this study. To briefly summarize the construction of this catalog, they employed SExtractor \citep{bertinSExtractorSoftwareSource1996} following a methodology similar to the one performed in the CANDELS catalog within the EGS field \citep{stefanonCANDELSMultiwavelengthCatalogs2017}; however, in this case,  the $F444W$-band served as the detection image. Fluxes and their respective errors for each source were measured using Kron elliptical apertures on images that were PSF-matched to the 0.16" FWHM angular resolution of the $F444W$-band images. For bands different from the $F444W$ detection band, they corrected to total fluxes and uncertainties using the ratio between Kron fluxes and isophotal fluxes as measured in $F444W$ by SExtractor. They applied an aperture correction for point sources in all bands to account for flux lost outside Kron's elliptical apertures and scaled the fluxes and uncertainties by the fraction of flux outside the Kron ellipses as measured in the $F444W$-band PSF. More details about the methodology can be found in \citet{gomez-guijarroJWSTCEERSProbes2023} and \citet{stefanonCANDELSMultiwavelengthCatalogs2017}.\\
\indent From this photometric catalog, \citet{gomez-guijarroJWSTCEERSProbes2023} estimated the photometric redshift for each source from a SED fitting procedure with the 18 fluxes and corresponding uncertainties from their catalog as input of the \texttt{EAZy-py} code, an updated Python version of the code \texttt{EAZy} \citep{brammerEAZYFastPublic2008}. To ensure the best photometric redshift determination possible, they employed a set of 13 Flexible stellar population synthesis \citep[FSPS;][]{conroyPropagationUncertaintiesStellar2010} templates \citep[$\rm corr\_sfhz\_13$;][]{kokorevALMALensingCluster2022} allowing the SED code to span various star-formation histories (SFHs), ages, and dust attenuation, while including the contributions of emission lines to the spectra. When possible thanks to  the spectroscopic redshift catalog of \citet{stefanonCANDELSMultiwavelengthCatalogs2017} or the MOSDEF survey of the EGS field \citep{kriekMOSFIREDeepEvolution2015}, photometric redshifts were replaced by spectroscopic redshifts. Here, we focus our analysis on the $3<z<4$ redshift range because our aim is to study the potential direct progenitors of QGs at cosmic noon. To ensure a good mass determination, we restricted our analysis to 1755 galaxies with photometric redshifts $3<z<4$ that are significantly detected ($>5\sigma$) in the $F444W$ and $F356W$-bands. They defined the parent sample of our study  which we have used to build our final and mass-complete samples.

\begin{figure}[h]
    \centering    \includegraphics[width=1\columnwidth]{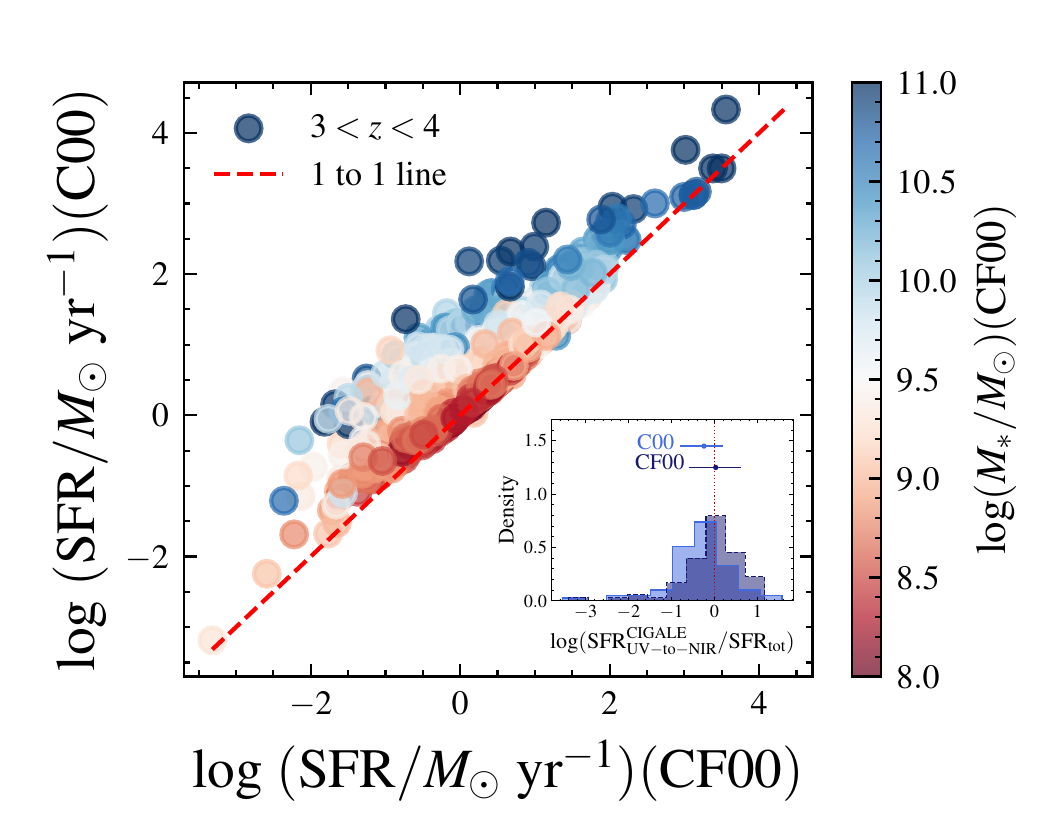}  
    \caption{Measurement comparisons of SFR obtained by SED-fitting with the CF00 and C00 attenuation law for the parent sample of 1755 galaxies at $3<z<4$ taken from the \citet{gomez-guijarroJWSTCEERSProbes2023} photometric catalog.  
    The inset shows the comparative histograms of the SFR ratio between the UV/NIR SFR obtained via an SED fitting with the CF00 (dark blue) and C00 (light blue) attenuation law and the $\rm SFR_{tot} = SFR_{UV}+SFR_{IR}$. The 
    UV/NIR data come from JADES and $\rm SFR_{tot}$ from the GOODS-ALMA 2.0 catalog.  
    For each distribution, the median value (filled circle) along with its associated $16\%$ and $84\%$ percentiles (segments) are displayed.  
    The red dotted vertical line indicates the benchmark whereby $100\%$ of the total SFR is retrieved by the UV/NIR SED fit.}
    \label{fig:why_CF00}
\end{figure}

\subsection{Mass, star formation rate, and attenuation law}
\label{Stellar mass, star formation rate and attenuation law}

\begin{table*}
\centering
\caption{\texttt{Cigale} model parameters}
\label{SEDsettings}
\begin{tabular}{l l l l}        
\hline\hline                 
Module & Parameter & Symbol & Values \\     
\hline                        
    \multirow{2}*{\makecell[l]{Star formation history \\ \textit{sfhdelayed}}} & $e$-folding timescale of the delayed SFH for the main population & $\tau_{\rm{main}}$ & 1-30 Gyr \\
     & Age of the main stellar population & $ t_{\rm{main}}$ & 0.1-3 Gyr\\
     \hline 
   \multirow{2}*{\makecell[l]{Simple stellar population \\ \textit{
   Bruzual and Charlot 2003}}} & Initial mass function & & Chabrier2003\\
    & Metallicity & $Z_{\odot}$ & 0.02\\
    \hline 
   \multirow{3}*{\makecell[l]{Dust attenuation \\ \textit{Charlot  Fall 2000}} }& V-band attenuation in the ISM & $A_{\rm V}^{\rm ISM}$ & 0-5\\
    & $\frac{A_{\rm V}^{\rm ISM}} {A_{\rm V}^{\rm BC}+A_{\rm V}^{\rm ISM}}$ & $\mu$ & 0.3\\ 
    &  Power-law slope of dust attenuation in the birth cloud & $n^{\rm BC}$ & $-0.7$ \\
    &  Power-law slope of dust attenuation in the interstellar medium & $n^{\rm ISM}$ & $-0.7$ \\
\hline                                   
\end{tabular}
\end{table*}

Galaxies of the parent sample were then fitted with \texttt{Cigale}. To compute the stellar component of our SED models, we assumed a delayed, exponentially declining star formation history (SFH) such as $\rm SFR  \propto \frac{1}{\tau^{2}}  \times exp(-t/\tau),$ where $\tau$ defines the time at which the SFR peaks, and used the \citep{bruzualStellarPopulationSynthesis2003} (BC03) stellar population synthesis model together with a \citet{chabrierGalacticStellarSubstellar2003} initial mass function (IMF) and the \citet{charlotSimpleModelAbsorption2000} (CF00) dust attenuation law (see below for more tests on that matter). To decide whether or not letting metallicity free during the fit, we investigated the constraining power of our UV/NIR data on this parameter by using the mock analysis from \texttt{Cigale} similarly to \citet{seilleSpatialDisconnectionStellar2022}. Briefly, \texttt{Cigale} uses the best-fit model fluxes and their associated observed photometric errors to run with the same configuration grid, but with flux values randomly chosen within a Gaussian distribution centered of the best-fit value and the photometric error as a standard deviation. \texttt{Cigale} then compares the result from the initial run with this run based on mock SED. Based on this analysis, we concluded that metallicity was not constrained with our data. Therefore, we decided to fix the metallicity to solar ($\rm Z=0.2$), as it has been shown to be a good assumption for massive ${\rm log}(M_\ast/M_\odot)> 9.6$ galaxies, according to the mass-metallicity relation at this redshift \citep{maOriginEvolutionGalaxy2016}. We also tested the contribution of a nebular continuum with emission lines on the properties derived from our mass-complete sample (defined in Sect.~\ref{Determination of the mass completeness}). We fit the galaxies with and without the nebular module and found no notable effects on the SFR, $A_{\rm v}$, and mass-weighted age. Nevertheless, we noticed a slight effect on stellar mass, with the inclusion of the nebular continuum underestimating the mass of ${\sim} 0.1$ dex on average compared to fit with the nebular component for ${\rm log}(M_\ast/M_\odot)<10$ galaxies, in agreement with \citet{yuanPropertiesLBGsOIII2019}. Based on these results, and since our main results are drawn from the most massive ${\rm log}(M_\ast/M_\odot)>10$ galaxies of our mass-complete sample, the choice of including nebular emission does not impact the results. Then, we decided to not include the nebular emission while fitting.

Since we are looking to study very dusty galaxies, the choice of an attenuation law is crucial. In fact, different laws can lead to different mass and SFR determinations if the dust emission is not considered, which is the case without FIR data \citep{buatColdDustStellar2019}. With \texttt{Cigale,} we can use a \citet{calzettiDustContentOpacity2000} (C00), or a \citet{charlotSimpleModelAbsorption2000} (CF00) attenuation-law. In C00, dust attenuation at a given wavelength is measured using the color excess as $A_{\lambda} = E(B-V)_{\rm stars}k_{\lambda}$ where $k_{\lambda}$ defines the reddening curve and $k_{\rm V}=\frac{A_{\rm V}}{ E(B-V)_{\rm stars}}=R_{\rm V}$. On the other hand, the CF00 assumes that young (<10 Myrs) and old stars do not undergo the same total attenuation. In this model, young stars have not yet left their birth clouds, so their light is attenuated by both the birth cloud and the ISM; whereas for old stars, their radiation is only affected by the dust in the ISM. To model these two attenuating components, CF00 considers two power-law attenuation curves, one for the birth cloud, $A_{\lambda}^{\rm BC} = A_{\rm V}^{\rm BC}(\lambda/0.55)^{n^{\rm BC}}$, and one for the ISM, $A_{\lambda}^{\rm ISM} = A_{\rm V}^{\rm ISM}(\lambda/0.55)^{n^{\rm ISM}}$, with the ratio between the attenuation of the two components controlled by $\mu=\frac{A_{\rm V}^{\rm ISM}}{(A_{\rm V}^{\rm BC}+A_{\rm V}^{\rm ISM})}$. Because the power-law indices and $\mu$ are not constrained by UV/NIR data \citep{buatGOODSHerschelDust2012,buatColdDustStellar2019,seilleSpatialDisconnectionStellar2022}, we opted for and then tested the initial CF00 formulation with $n^{\rm ISM}=n^{\rm BC}=-0.7$ and $\mu=0.3$. 

\begin{figure}[h]
\includegraphics{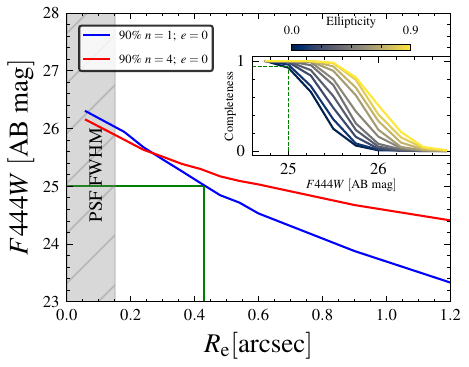}
\caption{Determination of the $F444W$ limiting magnitude. The dependence of the $90\%$ limiting magnitude on circularized effective radii from $0\farcs06$ to $1\farcs2$ is shown for an exponential profile ($n=1$, blue solid line) and a de Vaucouleurs profile ($n=4$, red solid line), both with null ellipticity ($e=0$). In the top right-hand corner, we show the completeness curves for a Sérsic profile $n=1$ with size $R_{\rm e}=0\farcs42$ and variable ellipticity and the projection of the $90\%$ completeness on the $e=0$ curve that is the most restrictive one. The  grey shaded region depicts the FWHM size of the $F444W$-band. The green solid lines defines the $F444W=25$ limiting magnitude as the projection on the most restrictive, while realistic, galaxy size ($R_{\rm e}=0\farcs42$) and ellipticity ($e=0$) at $z=3.5$.}  
\label{Completeness_determination}
\end{figure}

Firstly, we compared the SFRs inferred by \texttt{Cigale} while using the C00 or CF00 attenuation law on the parent sample. As the left panel of Fig.~\ref{fig:why_CF00} suggests, C00 and CF00 SFRs are similar for galaxies with low stellar mass (${\rm log}(M_\ast/M_\odot)<9.5$) at these redshifts, while high-mass sources present significantly higher SFRs when using CF00 attenuation law. We therefore investigated the accuracy of the inferred SFR of both attenuation laws for massive galaxies, taking advantage from the GOODS-ALMA 2.0 catalog \citep{gomez-guijarroGOODSALMASourceCatalog2022} that contains massive ALMA-detected sources in the Great Observatories Origins Deep Survey South field \citep[GOODS-South;][]{dickinsonGreatObservatoriesOrigins2003,giavaliscoGreatObservatoriesOrigins2004}. This catalog provides total SFRs that take into account unobscured and obscured SFRs, such that $\rm SFR_{tot} = SFR_{UV}+SFR_{IR}$. In this catalog, $\rm SFR_{IR}$ was obtained by using mid-IR to mm SED-fit of galaxies detected with a peaked significance level of $>3.5\sigma$ using ALMA Band 6. They derived the total $L_{\rm IR}$ by integrating the best-fit curve between rest frame 8 and 1000 $\mu$m, and subtracted a potential AGN contribution to IR emission. The deduced $L_{\rm IR}$ therefore only takes into account star formation, and is translated into $\rm SFR_{IR}$ using the \citet{kennicuttGlobalSchmidtLaw1998} conversion. Then, we cross-matched these 88 galaxies from the GOODS-ALMA 2.0 catalog, which are characterized by ${\rm log}(M_{\ast\rm median}/M_\odot)\sim 10.77$ and $z_{\rm median} =2.46$ to the JWST Advanced Deep Survey \citep[JADES;][]{eisensteinJADESOriginsField2023,eisensteinOverviewJWSTAdvanced2023} initial data release of the GOODS-South field presented in \citet{riekeJADESInitialData2023} to obtain their UV/NIR data. We used exactly the same nine photometric bands as those in our CEERS catalog to be as close as possible to the data used in our main study. These 88 galaxies are therefore analogs in terms of masses and redshifts to the $\rm RedSFGs$ defined in Sect.~\ref{Classifications} with ${\rm log}(M_{\rm \ast}/M_\odot)>10$. We performed SED fits on these galaxies and compared the SFRs obtained using UV/NIR data with a C00 or CF00 attenuation law to the total SFRs from the GOODS-ALMA 2.0 catalog. The right panel of Fig.~\ref{fig:why_CF00} suggests that the median SFR obtained by the CF00 law is in better agreement with the ALMA SFR, with $(\rm {SFR}_{\rm UV/NIR}^{\rm CIGALE}/SFR_{tot})_{median} = 1.1_{-4.2}^{+3.9}$ while the one from using the C00 law has $(\rm {SFR}_{\rm UV/NIR}^{\rm CIGALE}/SFR_{tot})_{median} = 0.57_{-3.7}^{+2.7}$. In view of this result, we decided to use the CF00 attenuation law. The \texttt{Cigale} grid used in the rest of the article (including the resolved SED-fitting) can be found in Table~\ref{SEDsettings}. It is important to note that the choice between the CF00 and C00 dust attenuation laws was not based on a comparison between the CEERS UV/NIR data in the EGS field and the FIR GOODS-ALMA data, but rather between the latter and the UV/NIR data from the JADES survey. Nevertheless, while CEERS images are shallower than JADES's in all NIRCam bands (e.g $F444W$ 5$\sigma$ point source depth of 29.77 in JADES and 28.56 in CEERS), these cross-matched galaxies are massive, so their SED fits do not suffer severely from the difference in depth. The conclusion would therefore hold if the same experiment were carried out in the CEERS field, but the absence of ALMA observations makes it impossible to reproduce.

\section{Determination and classification of the galaxies of the sample} 
\label{Determination and classification of the galaxies of the sample}

\subsection{Completeness}
\label{Completeness}

\subsubsection{Determination of the completeness in apparent magnitude}
\label{Determination of the completeness in apparent magnitude}

The completeness of a catalog is defined as the fraction of sources in a field that are actually detected as a function of their observed magnitudes. Nevertheless, the detectability of a source not only depends on its flux, but also on its surface brightness and, thus, on its morphology. As we aimed to study a mass-complete sample of galaxies at $3<z<4$ in this work, we evaluated the $90\%$ limiting magnitude in the $F444W$ detection band of our catalog for a large grid of profiles presented in Table~\ref{Injection-Recovery}.

We randomly injected 100 sources in each of the ten pointings in bins of magnitudes and with various Sérsic profiles across the whole field. We ensured that the injected profiles did not overlap with pre-existing sources as determined by the SExtractor segmentation map, and we avoided placing injected sources near the edges of the pointings or in the vicinity of a bright source. On these images, we ran the detection procedure used to generate the catalog presented in Sect.~\ref{Catalog} and computed the ratio $N_{\rm recovered}/N_{\rm injected}$ for each of the 10 fields. We subsequently calculated the mean of these values to generate the final completeness curves of the total field. The results are summarized in  Fig.~\ref{Completeness_determination} that shows how the $90\%$ limiting magnitude (i.e the magnitude at which $90\%$ of the sources are recovered) is influenced by the size and brightness profile of the source. It highlights that for galaxies with sizes close to the PSF FWHM of the detection band, the Sérsic index does not have a significant impact on the detection limit. Nevertheless, differences arise at $\rm {\sim} 2\times FWHM$ with higher Sérsic index (i.e., a cuspier surface brightness) profiles detected down to a higher magnitude than lower Sérsic index profiles. On the upper-right corner of this figure, we display different completeness curves of sources injected with ellipticities ranging from $e=0.9$ (highly elliptical) to $e=0$ (perfect circle) for a fixed $n=1$ and $R_{\rm e}=0\farcs42$. These choices of Sérsic index and  $R_{\rm e}$ to derive our detection limit are motivated by the broad morphological study of galaxies at redshift $z>3$ conducted in \citet{kartaltepeCEERSKeyPaper2023} using JWST/CEERS data in which they extracted Sérsic index and effective radius distributions for their sample. To derive a conservative estimate of the limiting magnitude, we used the results from the injection-recovery method of sources with $n=1$ together with the higher $R_{\rm e}$ found in \citet{kartaltepeCEERSKeyPaper2023} ($\rm 3kpc = 0\farcs42$ at $z=3.5$, the median redshift of our study), coupled with a null ellipticity. The orthogonal projection of this galaxy size on the $n=1$ curve with $e=0$ is shown as green solid lines.

\begin{table}
\centering
\caption{Simulation input parameters}
\label{Injection-Recovery}

\begin{tabular}{lll}        
\hline\hline                 
Parameter & &Values \\ 
\hline                        
$m_{4.4\mu \rm m}$[mag] & &20-30, step 0.25 \\
$n$ & &1,4 \\
$R_{\rm e}$[arcsec] & &0.06, 0.18, 0.24, 0.3, 0.36, 0.42, 0.48, \\
&&0.56, 0.6, 0.9, 1.2\\
$e=1-b/a$ & &0-0.9 step 0.1 \\
      
\hline
\end{tabular}
\end{table}

These results highlight that injection-recovering of only point sources will tend to artificially increase the limit in magnitude compared to extended sources, thus introducing potential bias into subsequent completeness-based analysis, which was already found in \citet{stefanonCANDELSMultiwavelengthCatalogs2017}. In the end, we determined a value of $m_{\rm lim,4.4\mu m}=25$ as the most conservative estimate for the limiting magnitude in the $F444W$-band.

\subsubsection{Final sample selection}
\label{Final sample selection}

Starting from the parent sample of 1755 galaxies, we carried out an analysis about photometric redshift uncertainties using different SED-fitting codes to verify the quality of our \texttt{EAzy}-based photometric redshifts. To this end, SED fitting on the parent sample was performed with \texttt{Beagle} \citep{chevallardModellingInterpretingSpectral2016}, \texttt{Cigale} \citep{boquienCIGALEPythonCode2019}, \texttt{Lephare} \citep{arnoutsMeasuringModellingRedshift1999,ilbertAccuratePhotometricRedshifts2006} and \texttt{HyperZ} \citep{bolzonellaPhotometricRedshiftsBased2000} assuming fitting grids as close as possible to the \texttt{EAZy} one. The median redshift values of these four SED codes were then compared with the redshifts determined by \texttt{EAZy}. The typical uncertainty on these photometric redshifts is $\sim 10\%$ and in the end, we imposed a relative error cut in the redshift determination, such that $\lvert z_{\rm EaZy}-z_{\rm median}\rvert/{z_{\rm EaZy}}<20\%$. Thus, in our redshift range of interest ($3<z<4$) and above the stellar mass limit defined in Sect.~\ref{Determination of the mass completeness}, the parent sample contains 237 galaxies. Of these, 21 ($\sim 9\%$) are excluded by this criterion, with only one that would have been identified as $\rm RedSFGs$ in Sect.~\ref{Classifications}. In what follows, we adopt the photometric redshifts from \texttt{EAZy}.

Next, we identified and removed galaxies with potential AGN contamination from our sample, as this population could distort the stellar properties of the resolved SED-fitting and, consequently, its interpretation. When looking for the signature of AGN with photometric data, we need to use different color criteria, as AGN can be of different natures and therefore detectable at different wavelengths. To seek for unobscured AGN, we cross-matched our sample to the AEGIS-XD catalog of \citet{luoCHANDRADEEPFIELDSOUTH2016} and computed X-ray luminosities in the rest-frame 2–10 keV band using the 0.5–10 keV flux $k$-corrected using a power-law X-ray spectrum with a spectral index of $\Gamma = 1.4$ similarly to \citet{kocevskiCEERSKeyPaper2023}. We then imposed the log($L_{\rm X}/L_\odot)>42.5$ threshold commonly used to differentiate unobscured AGN from star-forming galaxies \citep[e.g,][]{luoCHANDRADEEPFIELDSOUTH2016} and removed 8 out of 216 galaxies. For obscured AGN, we searched for any IR AGN contribution in the FIR catalog of Henry et al., (in preparation) that uses the super-deblending technique from \citet{liuSuperdeblendedDustEmission2018,jinSuperdeblendedDustEmission2018a}. We used the rest-frame UV/NIR fluxes from both our HST/ACS and JWST/NIRCam data and added the super-deblended photometry from Spitzer/IRAC 5.8, 8 $\rm \mu m$ and MIPS 24 $\rm \mu m$, Herschel/PACS 100, 160 $\rm \mu m$ and SPIRE 250, 350, 500 $\rm \mu m$ and JCMT/SCUBA2 450, 850 $\rm \mu m$. We performed a SED-fit using the Code Investigating GALaxy Emission \citep[\texttt{Cigale};][]{boquienCIGALEPythonCode2019} and fit the AGN contribution to the SED of galaxies with $\rm S/N>3$ in the 8 $\rm \mu m$ and 24 $\rm \mu m$-bands. By defining the AGN fraction, $f_{\rm AGN}$, as the AGN contribution to the total IR luminosity $L^{\rm AGN}_{\rm IR}=f_{\rm AGN} \times L^{\rm tot}_{\rm IR}$, we removed 4 out of 208 galaxies from our sample that have $f_{\rm AGN}>0.2$. Finally, as the accretion disk and the jets of AGNs produce synchrotron emission, radio detections have been historically used to detect heavily obscured AGNs. However, due to synchrotron emission from supernova remnants, SFGs can also be bright in the radio band, and this radiation correlates with their rest-frame 42 to 122 $\rm \mu m$ FIR emission \citep{helouThermalInfraredNonthermal1985,yunRadioPropertiesInfraredselected2001,magnelliFarinfraredRadioCorrelation2015} and (more generally) the rest-frame 8 to 1000 $\rm \mu m $ IR emission \citep{bellEstimatingStarFormation2003,delhaizeVLACOSMOSGHzLarge2017}. Therefore, a commonly used method is to identify AGN as galaxies that deviate by more than $3\sigma$ from the correlation followed by SFGs. Following \citet{delvecchioVLACOSMOSGHzLarge2017}, this is achieved by imposing a lower limit on the radio excess parameter, for instance, as in:
\begin{equation}
r={\rm log}(\frac{L_{1.4 \rm GHz}[\rm {W.Hz^{-1}}]}{\rm {SFR_{IR}}[\it {M}_{\odot}.\rm {yr^{-1}}]})>22~(1+z)^{0.013} 
,\end{equation}
where $L_{1.4 \rm GHz}$ is the 1.4 GHz rest-frame luminosity and $\rm SFR_{IR}$ is computed from the integrated 8–1000 $\rm \mu m $ IR luminosities. Here, we use the super-deblended FIR catalog of Henry et al., (in preparation) that provides $\rm SFR_{IR}$ from FIR+submm (observed 100 $\mu$m to 1.1 mm) SED fit and $L_{1.4 \rm GHz}$ using the $\rm 3~GHz$ flux density following \citet{wangConstraintsCosmicStar2024}. We removed 1 out of 204 additional source following this method. 
We also report the presence of five sources with extremely low luminosity given their stellar mass of ${\rm log}(M_\ast/M_\odot)\sim 10$ in our mass-complete sample. Visual inspection of these sources revealed their extremely small size while showing an extremely high dust attenuation ($A_{\rm v, median}^{\rm ISM}\sim 4$) and SED expected for "little red dots" \citep{labbeUNCOVERCandidateRed2023a,kokorevCensusPhotometricallySelected2024}. Indeed, three of them are selected as LRD by the SED's continuum slope fit from \citet{kocevskiRiseFaintRed2024} with two being spectroscopically confirmed $z_{\rm spec}=5.08$ and $z_{\rm spec}=5.62$ highly obscured AGN showing broad emission lines with NIRSpec (CEERS 5760 in \citet{kocevskiRiseFaintRed2024} and CEERS 82815 in \citet{kocevskiHiddenLittleMonsters2023}). We therefore decided to exclude them from this analysis.
In the end, we removed 18 potential AGN contaminants from the mass-complete sample of 216 galaxies with secure photometric redshifts.

\begin{figure}
\includegraphics{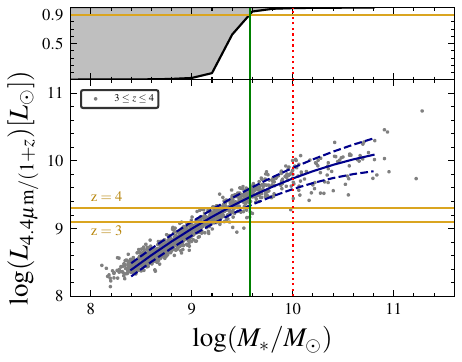}
\caption{$\textit{Lower panel}$: Distribution in stellar mass and luminosity at rest-frame wavelength $4.4\mu {\rm m}/(1+z)$ with $3 \leq z \leq 4$ for the final sample of 1202 galaxies. The best-fit relation with its $1\sigma$ dispersion are displayed in solid and dashed blue lines, respectively. The position of the $F444W= 25$ limiting magnitudes for $z = 3$ and $z = 4$ are shown with two golden lines. $\textit{Upper panel}$: Evolution of completeness with stellar mass at the $z=4$ luminosity limit. The horizontal golden line showing the $90\%$ completeness level. Consequently, the green vertical line defines the estimated $90\%$ mass completeness limit. 
Taking into account galaxies with low mass-to-light ratios, our sample is $99\%$ complete for this population to the right of the red dotted line.}  
\label{fig: completness_limZ_relativeerrormedian9}
\end{figure}

Finally, to ensure reasonable parameter estimates, we excluded poorly fitted galaxies, defined as galaxies with $\rm \chi_{reduced}^{2}>1.68$ as this upper limit corresponds to a significance level of $10\%$ for data with 8 degrees of freedom (9-1 photometric bands).In other words, a significance level of $10\%$ implies that there is a $10\%$ probability that the observed $\rm \chi_{reduced}^{2}$ value would be as large or larger purely by chance if the model were correct. Therefore, galaxies with $\rm \chi_{reduced}^{2}$ values above this limit are considered to have unacceptable fits, indicating that the model does not adequately describe the observed photometry for these galaxies. We removed 10 out of 198 additional galaxies based of this criteria in our mass-complete sample derived in Sect.~\ref{Determination of the mass completeness}. In the end, the final sample consists of 1202 galaxies, free from uncertain redshifts, potential AGN contaminants, and poorly fitted objects. 

\subsection{Determination of the mass completeness}
\label{Determination of the mass completeness}

We used the final sample defined above to define the mass completeness and, ultimately, the mass-complete sample that constitutes the working sample of this study.

In order to obtain a mass-complete sample of galaxies detected at $3<z<4$, the apparent limiting magnitude must be translated into a limiting mass above which we control the fraction of galaxies that could have been missed by our detection method. However, mass determination suffers from systematic uncertainties even when NIR data are available as its determination can be affected by the SED-fitting code used, the prior assumptions made to compute the stellar populations \citep{gawiserSpectralEnergyDistribution2009,lowerHowWellCan2020,pacificiArtMeasuringPhysical2023,narayananOutshiningRecentStar2024}, the redshift uncertainty, and most importantly the intrinsic variation of the SED shape of each galaxy according to its formation and evolution history. This in turn implies that it is very inaccurate to simply convert the apparent magnitude limit determined in Sect.~\ref{Determination of the completeness in apparent magnitude} into a limiting mass using simply a constant stellar mass-to-light ratio.

We therefore determined the $90\%$ mass-completeness limit empirically following \citet{schreiberHerschelViewDominant2015}, and exploited the relation between the rest-frame $L^{rest}_{\lambda}$ where $\lambda=4.4/(1+z) = 1\pm 0.1 ~\rm {\mu m}$ in this range of redshift, and the stellar mass inferred by \texttt{Cigale} using the full UV/NIR photometry of each galaxy. As shown in Fig.~\ref{fig: completness_limZ_relativeerrormedian9}, the mass versus light relation and its corresponding $1\sigma$ scatter vary with mass, the later ranging from $\rm {\sim} 0.18$ dex at low mass (i.e., ${\rm log}(M_\ast/M_\odot)\sim 8.4$) to $\rm {\sim} 0.58$ dex at high mass (i.e., ${\rm log}(M_\ast/M_\odot)\sim 11$). 

\begin{figure}[h]
\centering
\includegraphics[width=0.5\textwidth]{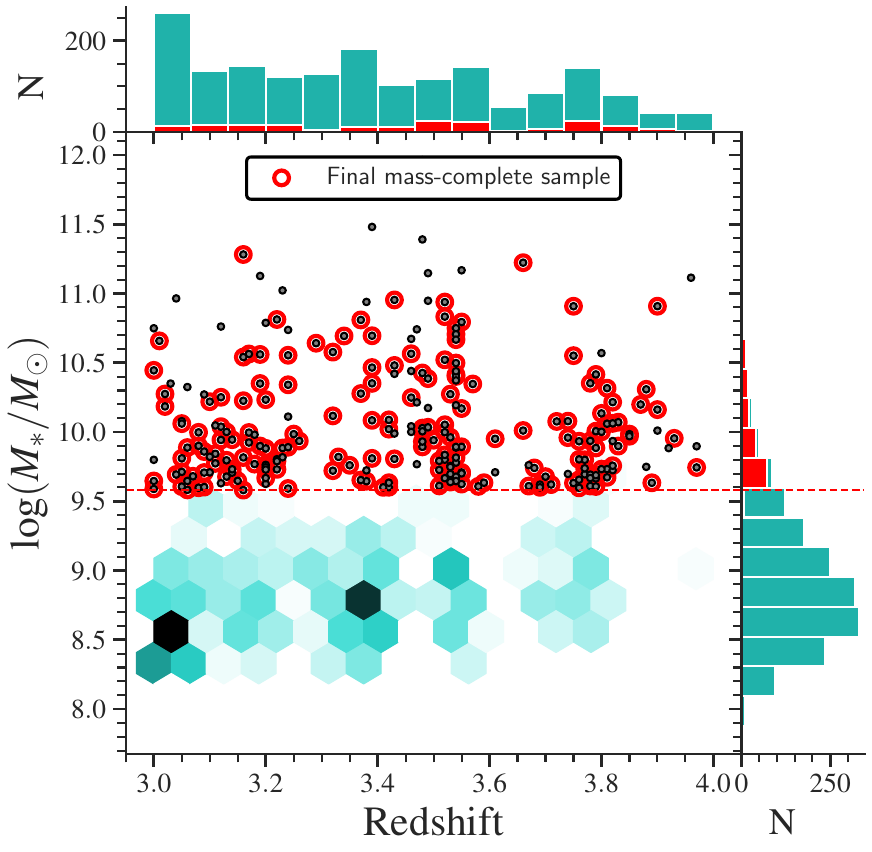}  
\caption{Distribution of the parent and final mass-complete samples in terms of stellar mass and redshift over the studied redshift range. The parent sample is represented as a light green-scale density map. The horizontal red dashed line indicates the limiting mass of ${\rm log}(M_\ast/M_\odot)>9.6$ where the parent sample is $90\%$ complete and contains 237 galaxies. Sources encircled in red define the final mass-complete sample of 188 galaxies used in this paper. This sample is clean from objects with uncertain redshifts, potential AGN contaminants or bad SED fits. The redshift and stellar mass distributions of the two samples are shown as marginal histograms of their respective colors.}  
\label{fig: presentation_sample}
\end{figure}

This increasing scatter is mainly due to the presence of a population of galaxies with ${\rm log}(M_\ast/M_\odot)> 10$ that are outliers of the main relation by showing high mass-to-light ratio. It appears that all of these galaxies that deviate by more than $1\sigma$ from the main relation are classified as $\rm RedSFGs$ in Sect.~\ref{Classifications}, therefore hinting at their highly dusty nature ($A_{\rm v, median}^{\rm ISM}\sim 2.2$). We incorporated the evolution of the mass-to-light ratio and its scatter in our limiting mass determination by fitting them with a second degree polynomial function. Using the best fit equations, we generated Gaussian distributions of mock galaxies in stellar mass bins and computed the fraction of galaxies that have their luminosities greater than a given limiting luminosity. Because a source with a given flux can have different deduced rest-frame luminosities depending on its redshift, to be as conservative as possible, we imposed the $z=4$ limiting luminosity and considered our sample to be complete in mass when completeness exceeded $90\%$. 

The final mass-complete sample comprises 188 galaxies above a limiting mass of ${\rm log}(M_{\rm \ast lim}/M_\odot)=9.6$ and is cleaned of uncertain redshifts, bad fits, and AGN contaminants (see Sect.~\ref{Final sample selection}). It is shown with its original parent sample in Fig.~\ref{fig: presentation_sample}. We note that although this $90\%$ completeness threshold is commonly used in the literature, the $10\%$ of galaxies missed down to this mass limit have by definition a particularly high mass-to-light ratio and are in practical associated with $\rm RedSFGs$ defined in the following section. This population constitutes the focus of this study and could be significantly biased against lower mass $\rm RedSFGs$ if we consider this mass limit. We therefore need to impose an even stricter completeness threshold of $99\%$, which is reached at ${\rm log}(M_{\rm \ast}/M_\odot)=10$ when comparing galaxies' properties with those of $\rm RedSFGs$. Thereafter, we qualify the sample with ${\rm log}(M_{\rm \ast}/M_\odot)>9.6$ as complete in mass and we specify when we are comparing subpopulations of galaxies with ${\rm log}(M_{\rm \ast}/M_\odot)>10 $ in particular.

\subsection{Classifications}
\label{Classifications}

The goal of this study is to examine the connection between $\rm BlueSFGs$, $\rm RedSFGs,$ and QGs. To achieve this, the first step was to categorize each galaxy as SFG or QG. We identified the QG population among our mass-complete sample of 188 galaxies by using the union of the widely used \textit{UVJ} color criteria \citep{patelUVJSelectionQuiescent2012,fumagalliHowDeadAre2014,fangDemographicsStarformingGalaxies2018,valentinoQuiescentGalaxiesBillion2020,valentinoAtlasColorselectedQuiescent2023,carnallMassiveQuiescentGalaxy2023} with a criteria based on the distance below the MS: $\rm \Delta MS = SFR/SFR_{MS}$ where the SFR in computed by \texttt{Cigale} and is compared to the MS from \citet{schreiberHerschelViewDominant2015}. Here, we used the \textit{UVJ} diagram as defined by \citet{whitakerNEWFIRMMEDIUMBANDSURVEY2011} : \\ 
\[
\text{Quiescent} = \left\{
    \begin{array}{lll}  
    U-V > 1.2, \\
    V-J < 1.4, \\
    U-V > 0.88 \times (V-J) + 0.59, \\
    \end{array}
\right.
\]
and defined QGs, sources that enter the \textit{UVJ} color selection or have $\rm \Delta MS < 0.6$ ($ \rm {\sim} 1.2$ dex below the MS). We checked for their IR detection in the catalog of Henry et al. in preparation and verified that these sources were not significantly detected ($\rm SNR\sim1$) in observed bands redder than $24\mu$m. While two of them have $\rm SNR\sim2$ in the $24\mu$m band, they all have $\rm{SFR}<100 M_\odot~\rm{yr^{-1}}$, placing them below the 0.5 dex scatter of the MS. This way, we ensured that our QG selection selects objects that have sufficiently evolved to be treated apart from SFGs.

Secondly, we separated the population of red, dust-obscured SFGs (i.e., $\rm RedSFGs$) from the blue SFGs (i.e., $\rm BlueSFGs$) by identifying SFGs particularly faint in the rest-frame UV/optical wavelengths. In our redshift range, this intrinsic faintness translates into a faint/extremely faint detection in the optical-to-NIR up to and including the $H$-band. Previous selection criteria \citep{wangDominantPopulationOptically2019,xiaoHiddenSideCosmic2023} were based on HST/WFC3 images in the H-band that were shallower than JWST/NIRCam images in the $F150W$-band. Here, NIRCam data allow us to go a deeper and study sources that are fainter in the observed optical. We therefore decided to adopt  \\
\[
\text{$\rm RedSFGs$} = \left\{
    \begin{array}{ll}  
    F150W - F444W > 2, \\
    F444W < 25. \\
    \end{array}
\right.
\]

\begin{figure}[h]
\includegraphics{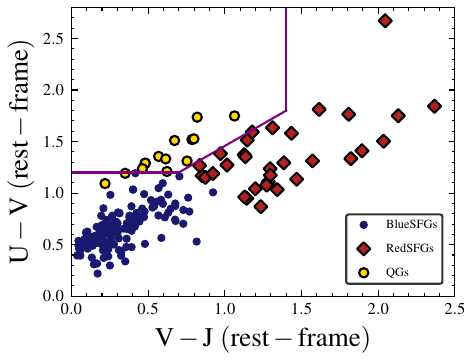}
\caption{Location of $\rm BlueSFGs$, $\rm RedSFGs$ and QGs at $3<z<4$ in the \textit{UVJ} color-color diagram, the purple line delimits the quiescent region as defined in \citet{whitakerNEWFIRMMEDIUMBANDSURVEY2011}.}  
\label{fig: cleaned_UVJ}
\end{figure}

\begin{figure}[h]
\includegraphics{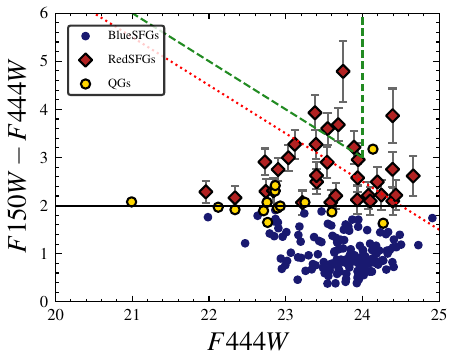}
\caption{$F150W-F444W$ versus $F444W$ color-magnitude distribution for our $3<z<4$ sample. The markers show the color-selected populations of $\rm BlueSFGs$, $\rm RedSFGs$ and QGs located above the black solid line defining our $\rm RedSFGs$ selection criteria. As a reference, we indicate the selection criteria for $H$-dropouts in \citet{wangDominantPopulationOptically2019} ( i.e., $F150W > 27$ mag and $F444W < 24$ mag) as a green dashed triangular region and the OFG selection from \citet{xiaoHiddenSideCosmic2023} (i.e., $F150W > 26.5$ mag and $F444W < 25$ mag) as a red dotted line.}  
\label{cleaned_OFG_selection.pdf}
\end{figure}

Galaxies that do not fit the QGs and $\rm RedSFGs$ criteria are then labeled $\rm BlueSFGs$. They define the population of more "typical" SFGs and our control sample. The population of $\rm RedSFGs$ described here includes dusty galaxies with strong far-IR emission, generally referred to as dusty star-forming galaxies (DSFGs). However, the color selection made here does not guarantee that all of them are, so we prefer to keep the name $\rm RedSFGs$ to avoid possible confusion. We can see, when looking at the distribution of our sample in the \textit{UVJ} diagram in Fig.~\ref{fig: cleaned_UVJ}, that our $\rm RedSFGs$ criteria selects galaxies that are located apart from the locus of $\rm BlueSFGs$ and preferentially in the DSFG region. In Fig. \ref{cleaned_OFG_selection.pdf}, the $H$-dropout selection criteria of \citet{wangDominantPopulationOptically2019} is shown by the green dotted triangular region and the \citet{xiaoHiddenSideCosmic2023} color selection is represented with a red dotted line. The black solid line marks the $\rm \Delta MS$ criteria used in this study. The $\rm RedSFGs$ are shown as red diamonds and lie above this delimiting line. By these criteria, we classified 32 $\rm RedSFGs$ (${\sim} 17\%$), 142  $\rm BlueSFGs$ (${\sim}76\%$) and 14 QGs (${\sim} 7\%$) in our cleaned mass complete sample of 188 galaxies with $3<z<4$ defined in Sect.~\ref{Determination of the mass completeness}. We recall that between ${\rm log}(M_{\rm \ast}/M_\odot)=9.6$ and ${\rm log}(M_{\rm \ast}/M_\odot)=10$, our sample can be considered complete for QGs and $\rm BlueSFGs$ but incomplete for $\rm RedSFGs$. Above ${\rm log}(M_{\rm \ast}/M_\odot)=10$, we can consider our sample complete for $\rm BlueSFGs$, $\rm RedSFGs$ and QGs.

\section{Results}
\label{Results}

\subsection{Integrated properties}
\label{Integrated Properties}

\begin{figure}
\includegraphics{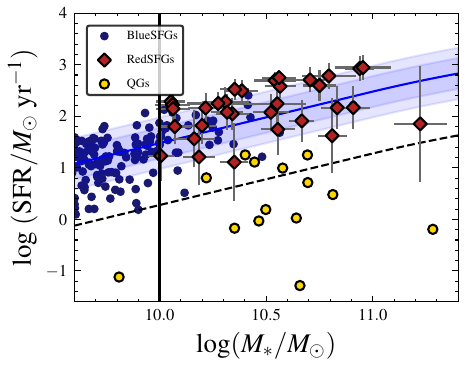}
\caption{Location of $\rm BlueSFGs$, $\rm RedSFGs$, and QGs at $3<z<4$ in the ${\rm SFR}-M_\ast$ plane. The \citet{schreiberHerschelViewDominant2015} MS is displayed as a solid blue line. We represent the $1 \sigma$ scatter of this MS associated with $ \rm 0.5 < \Delta MS < 2 ~~({\sim} 0.3~dex)$ as a shaded blue area as well as a wider scatter of $ \rm 0.33 < \Delta MS < 3 ~~({\sim}0.5~dex)$ in lighter blue. The $\rm \Delta MS<0.6~dex$ criterion used in combination with the UVJ diagram to define QGs. We rescaled all SFR values to the median redshift of our mass-complete sample ($z_{\rm median} = 3.51$) as explained in the main text. The sample is mass-complete for the three populations above the thick black line.} 
\label{cleaned_MS.pdf}
\end{figure}

\begin{figure}
\includegraphics{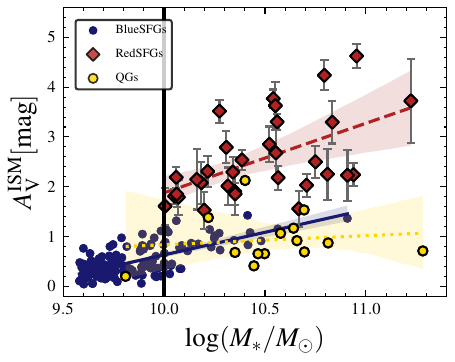}
\caption{Dust attenuation as a function of stellar mass for our final $3<z<4$ sample. $\rm BlueSFGs$, $\rm RedSFGs$ and QGs are represented with blue dots, red diamonds and yellow dots with black circle, respectively. Their regression lines and ${\sim} 95 \%$ uncertainties are shown with a solid, dashed and doted line with a shaded area. We indicate the $99\%$ mass-completeness limit of the sample with a thick black line.}  
\label{cleaned_AvMassbayes.pdf}
\end{figure}

Before studying the resolved properties of $\rm BlueSFGs$, $\rm RedSFGs,$ and QGs, we compared their integrated properties such as their stellar mass, SFR, and dust attenuation. In Fig.~\ref{cleaned_MS.pdf}, we show the star-forming MS arising from our mass-complete sample. Here, the SFRs are rescaled to the median redshift of the mass-complete sample: $z_{\rm median}=3.49$. This is done by conserving $\Delta \rm MS$ for each galaxy relative to the MS associated with their respective redshift. In this way, the rescaled SFRs represent the overall distribution of the sample in the SFR-stellar mass plane and this ensures that observed differences in SFRs are not simply due to the evolving MS between $z=3$ and $z=4$ but reflect intrinsic differences in galaxy properties. As expected, the majority of color-selected SFGs lie on the MS within the $1 \sigma$ uncertainty. In particular, we note that $\rm RedSFGs$ also fall on the MS, although, the lack of mid-IR-to-mm data that are necessary to constrain the SFR of highly dusty galaxies \citep{buatColdDustStellar2019,figueiraSFREstimationsComparison2022} may induce an  underestimation of the SFR for some of them. The location of the QGs in the SFR-stellar mass plane, below the MS, confirms that the redness of these galaxies is not due to dust attenuation as is the case for $\rm RedSFGs$, but rather to their old stellar population. This is also illustrated in Fig.~\ref{cleaned_AvMassbayes.pdf} where QGs are not as dusty ($A_{\rm v,median}^{\rm ISM}\sim 0.8$) as $\rm RedSFGs$ ($A_{\rm v,median}^{\rm ISM}\sim 2.2$) despite their similar stellar mass (${\rm log}(M_{\ast \rm median} /M_\odot)\sim 10.5$. We note here that despite the higher stellar mass of QGs compared to $\rm BlueSFGs$ (${\rm log}(M_{\ast \rm median}/M_\odot)\sim 9.8$), their dust attenuation does not evolve with mass -- as is the case for $\rm BlueSFGs$ and $\rm RedSFGs$. They display dust attenuation similar to that of $\rm BlueSFGs$ with ${\rm log}(M_\ast/M_\odot)\sim 10.2$. This low $A_{\rm v}^{\rm ISM}$ demonstrates that despite being compact and cuspy (see Sect.~\ref{Results_radial_profiles}), these galaxies have stopped funneling gas and dust in their cores, which would have allowed star formation to continue. Instead, they seem to have reached a physical stability. We also witness a clear offset in dust attenuation between $\rm BlueSFGs$ and $\rm RedSFGs$, with the latter being significantly more attenuated than the first at all stellar masses. Because such a distinction could be due to a bimodal distribution of ellipticities (edge-on galaxies tend to be more attenuated than face-on ones, as the column density of dust and gas along the line of sight is greater), we measured the median ellipticity of $\rm RedSFGs$ on the $F444W$ images. We found the median ellipticity of $\rm RedSFGs$ ($e_{\rm median}= 0.5 \pm 0.17$) to be similar to the one of ${\rm log}(M_\ast/M_\odot)> 10$ typical $\rm BlueSFGs$ ($e_{\rm median}= 0.52 \pm 0.17$), but no trend was found between their ellipticities and dust-obscuration in both cases. This significantly high dust obscuration in $\rm RedSFGs$ is  not therefore attributed to a geometrical effect. 

\begin{figure}
\includegraphics{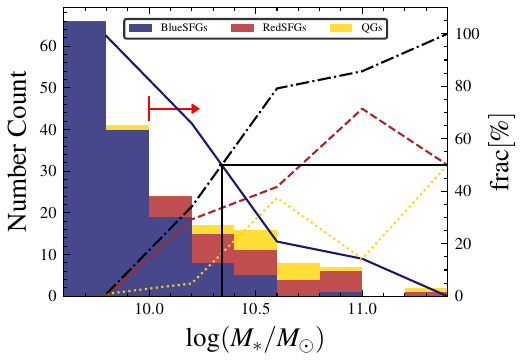}
\caption{Mass distribution and relative fractions of the 142 $\rm BlueSFGs$ (blue solid line), 32 $\rm RedSFGs$ (red dashed line), and 14 QGs (yellow dotted line) as a function of stellar mass. The cumulative contribution of $\rm RedSFGs$ and QGs to the total number count as a function of stellar mass is displayed as a black dash-dotted line. The black solid lines show the stellar mass above which $>50\%$ of galaxies are either $\rm RedSFGs$ or QGs.  The arrow indicates the mass completeness limit for the $\rm RedSFGs$ population.} \label{histogram_Mass.pdf}
\end{figure}

We  also studied the relative fractions of $\rm BlueSFGs$, $\rm RedSFGs,$ and QGs as a function of stellar mass and indicated them with solid and dotted colored lines in Fig.~\ref{histogram_Mass.pdf}. Here, we observe a decrease in the $\rm BlueSFGs$ fraction with increasing stellar mass and an opposite behavior for $\rm RedSFGs$ and QGs that becomes more and more dominant at high mass. At ${\rm log}(M_\ast/M_\odot)> 10.4$, the $\rm BlueSFGs$ fraction drops below $50\%$, marking the end of their domination. Consequently, QGs and particularly $\rm RedSFGs$ dominate for ${\rm log}(M_\ast/M_\odot)> 10.4$. Above ${\rm log}(M_\ast/M_\odot)> 10.6$, $\rm RedSFGs$ account for over $50\%$ of all galaxies, making them the dominant high-mass population in this redshift range. We note that because of $\rm RedSFG$ incompleteness for ${\rm log}(M_\ast/M_\odot)< 10$, their fractions might be underestimated in this mass regime. Finally, we find that the number density of the $\rm RedSFG$ sample $n\sim 1.1 \times 10^{-4}~\rm Mpc^{-3}$ is comparable to the one of compact ($R_{\rm e}\sim 1$ kpc) and massive ${\rm log}(M_\ast/M_\odot)> 10.5$ QGs at $z\sim 2$ \citep{vanderwel3DHSTCANDELSEvolution2014,vandokkumFormingCompactMassive2015}. This is an indication that this population could be the direct progenitor of these QGs found at cosmic noon. In the following, we investigate this $\rm RedSFG$ population and its role in the evolution of SFGs toward quiescence by comparing its resolved properties to those of $\rm RedSFGs$ and QGs.

\subsection{Methodology for obtaining the radial profiles }
\label{Radial Profiles Methodology}

\begin{figure*}[h]
\centering
     \includegraphics[width=0.9\textwidth]{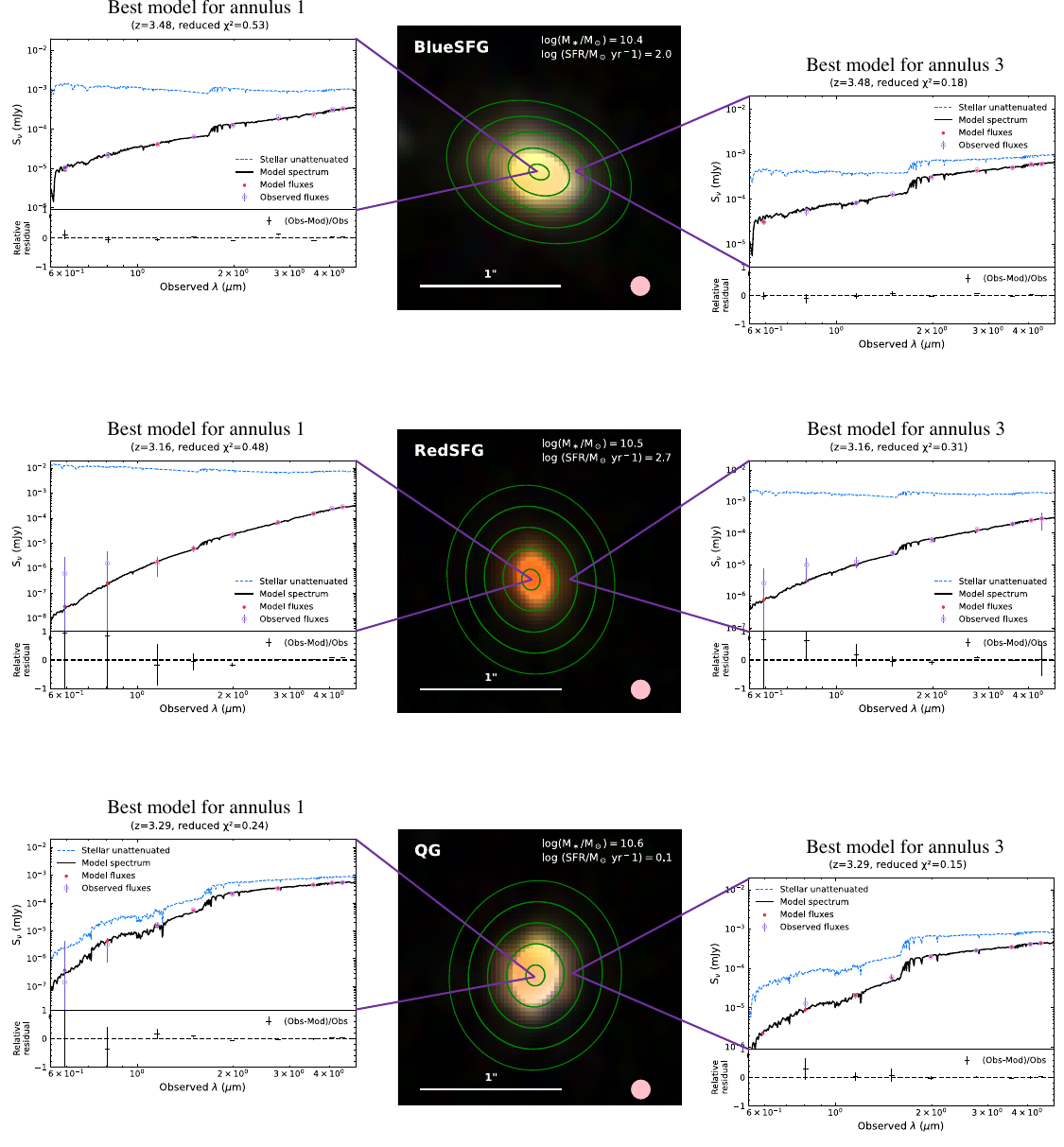}
      \caption{Illustration of the methodology used to compute radial profiles. From top to bottom, we show an example of $\rm BlueSFG$, $\rm RedSFG,$ and QG RGB cutouts generated using the $F200W$, $F277W$, $F356W$-bands PSF-matched to the $F444W$ filter. Each cutout has a $2\farcs{\rm \times 2}\farcs$ angular size and contains a $1\farcs$ white bar that defines the scale of the image. The $0.16\farcs$ angular resolution of the $F444W$-band is represented by a pink circle. Its radius corresponds to the semi-major axis of annulus 1 as well as the width of each annulus. For each galaxy, five concentric rings with a constant width equal to the FWHM of the $F444W$-band are displayed. In addition, for each cutout, we present the best model spectra produced by \texttt{Cigale} for annulus 1 (the innermost) and annulus 3. In those, the blue curve is the unattenuated stellar emission}
       
\label{test_galaxie.pdf}
\end{figure*}

To obtain spatially resolved stellar population properties for each galaxy in our mass-complete sample, maintaining a high signal-to-noise ratio (S/N) of typically $\rm S/N > 3$ in the fitted regions was key. This is especially important for bands probing the Balmer break, essential for reliably measuring dust obscuration. Due to higher dust extinction in $\rm RedSFGs,$ compared to $\rm BlueSFGs$, neither simple pixel-by-pixel binning nor Voronoi binning \citep{cappellariAdaptiveSpatialBinning2003} could ensure trustworthy S/N in the UV/optical-bands, resulting in an unfair comparison between the two SFG populations. Consequently, we opted to bin each galaxy into concentric elliptical annuli centered on the brightest pixel value in the $F444W$-band, with a width corresponding to the size of the $F444W$ PSF ($0\farcs16$ FWHM). The position angles and ellipticities used for each galaxy were obtained by fitting the corresponding $F444W$ segmentation maps with an ellipse using \texttt{Photutils} \citep{larry_bradley_2024_12585239}. Prior to that, we PSF-matched all images to the $F444W$ PSF. We show in Fig.~\ref{test_galaxie.pdf} our binning method on RGB, PSF-matched images of a typical $\rm BlueSFG$, an $\rm RedSFG$ and a QG. This method allowed us to have $\rm S/N > 3$ in each annulus on the $F150W$ images.

We computed the total flux in each of the ten photometric bands for each annulus by summing the fluxes on a pixel-by-pixel basis. For the total errors in a given annulus and in a given band, we measured total flux within 20 elliptical apertures in the vicinity of the sources that contained the same number of pixels as the annulus under consideration, and calculated the standard deviation of these measured fluxes. The errors were then assumed to be constant for all pixels belonging to an annulus. Finally, we conducted SED-fitting for each annulus using \texttt{Cigale} using the same grid as in Sect.~ \ref{Stellar mass, star formation rate and attenuation law} and shown in Table \ref{SEDsettings}. We computed the stellar mass density ($\rm \Sigma_{\star}$) and the star formation rate density ($\rm \Sigma_{SFR}$) by dividing the total mass and SFR in the annulus by the total area  within each of them.

Figure \ref{test_galaxie.pdf} shows for each studied sub-population, the best fit output by \texttt{Cigale} for the innermost annulus and the third outer annulus (referred to as annulus 1 and 3, respectively). These images and best fits highlight the peculiarity of $\rm RedSFGs$ as being more compact and redder than $\rm BlueSFGs$ and QGs for a given stellar mass. We also acknowledge the high dust obscuration of $\rm RedSFGs$ that significantly impacts the UV and optical part of their SED. 

To compare the profiles of the three populations, which span a wide range of stellar mass and SFR, it is necessary to apply a re-normalization in order to remove any dependence of mass and SFR on radial profiles. For example, it is well-known that SFGs follow a mass-size relation \citep{kormendyProposedRevisionHubble1996,shenSizeDistributionGalaxies2003,trujilloSizeEvolutionGalaxies2006,buitragoSizeEvolutionMost2008,bruceMorphologiesMassiveGalaxies2012,onoEVOLUTIONSIZESGALAXIES2013,vanderwel3DHSTCANDELSEvolution2014,langeGalaxyMassAssembly2015,allenSizeEvolutionStarforming2017,dimauroStructuralPropertiesClassical2019,mowlaCOSMOSDASHEvolutionGalaxy2019,nedkovaExtendingEvolutionStellar2021,wardEvolutionSizeMass2024}, with the most massive SFGs being more extended, whereas QGs exhibit a distinct sequence in the mass-size plane \citep{wardEvolutionSizeMass2024} and $\rm RedSFGs$ may also adhere to a different relation due to their pronounced compactness at higher masses compared to $\rm BlueSFGs$ \citep{gomez-guijarroJWSTCEERSProbes2023}. We took this mass effect into account by rescaling the radii to the median mass of our $\rm BlueSFGs$ sample (${\rm log}(M_{\ast \rm median}/M_\odot)\sim 9.8$) using the \citet{wardEvolutionSizeMass2024} mass-size relation of SFGs. Similarly, we re-scaled the $\rm \Sigma_{\star}$, $\rm \Sigma_{SFR}$ and $A_{\rm v}^{\rm ISM}$ profiles to this same median mass using the \citet{schreiberHerschelViewDominant2015} MS at $z_{\rm median}=3.49$ and the $A_{\rm v}^{\rm ISM}-{\rm log}(M_\ast/M_\odot)$ linear relation of $\rm BlueSFGs$ shown in Fig.~\ref{cleaned_AvMassbayes.pdf}. This way, we ensured a fair comparison between these three galaxy populations, allowing us to place the QGs and $\rm RedSFGs$ in the context of $\rm BlueSFGs$. To compare their morphologies, we computed the median of these re-normalized radial profiles for $\rm \Sigma_{\star}$, $\rm \Sigma_{SFR}$, $A_{\rm v}^{\rm ISM}$, sSFR, and mass-weighted age.

\begin{figure*}[h!]
\centering
     \includegraphics[width=1\textwidth]{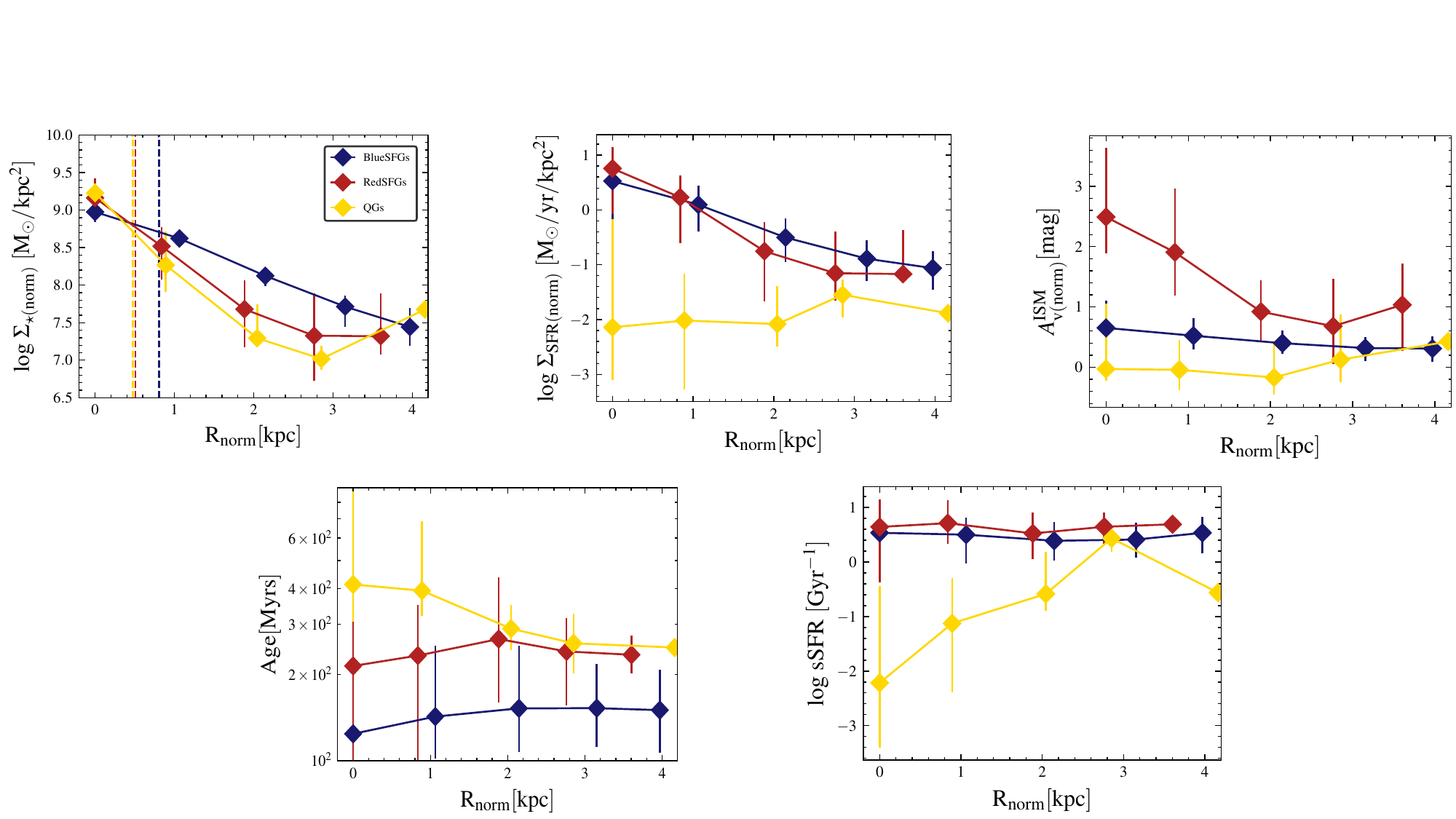}
      \caption{From left to right and top to bottom: Stellar mass density ($\rm \Sigma_{\star}$), star formation rate density ($\rm \Sigma_{SFR}$), dust attenuation ($A_{\rm v}^{\rm ISM}$), sSFR, and mass-weighted age normalized median radial profiles inferred using \texttt{Cigale} for $\rm BlueSFGs$ (in blue), $\rm RedSFGs$ (in red), and QGs (in yellow). Error bars represent the 68th and 32th quantiles of the distributions in each radial bin. In the $\rm \Sigma_{\star}$ panel we indicate the (normalized) half-mass effective radius for each population with a vertical dashed line.}
\label{Radial_profiles}
\end{figure*}

\begin{figure*}[h]
\centering
     \includegraphics[width=1\textwidth]{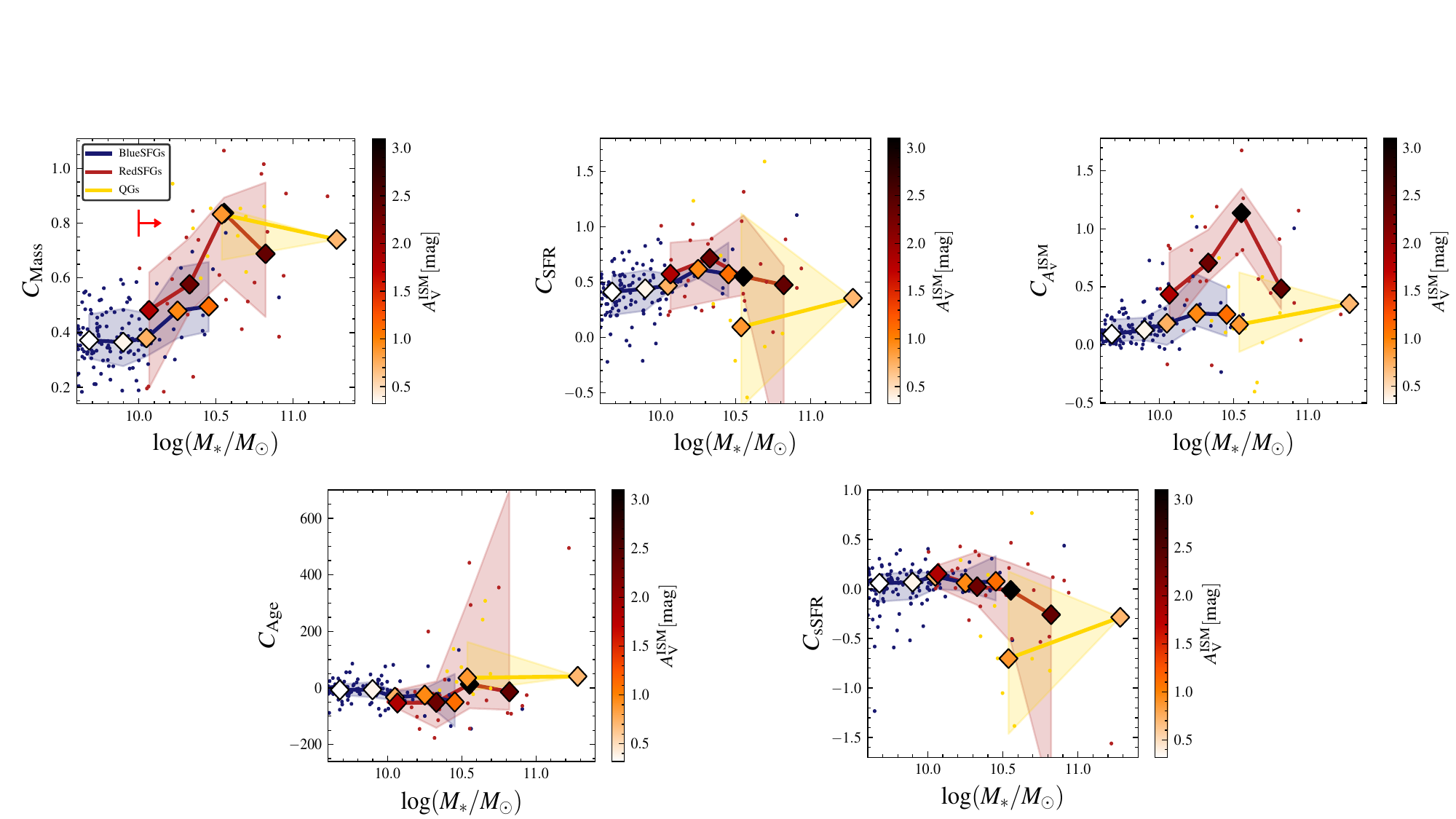}
      \caption{From left to right and top to bottom: $\rm \Sigma_{\star}$, $\rm \Sigma_{SFR}$, $A_{\rm v}^{\rm ISM}$, sSFR, and mass-weighted age concentrations as a function of stellar mass. The different concentrations are determined by measuring the slopes of the three innermost annuli of each curve individually. In each panel, $\rm BlueSFGs$, $\rm RedSFGs$, and QG are marked by blue, red and yellow dots, respectively. The medians in stellar mass bins are displayed for each population as diamonds, color-coded by the median dust attenuation in the bin. Uncertainties are represented by shaded regions of the color of the represented population and represent the 68th and 32th quantiles of the distributions in each stellar mass bin. The medians are centered on the median stellar mass of each bin containing at least four galaxies (with the exception of the most massive QG bin, which contains only one galaxy)}.
\label{Slopes}
\end{figure*}

\subsection{Radial profiles in $\rm \Sigma_{\star}$, $\rm \Sigma_{SFR}$, $A_{\rm v}^{\rm ISM}$, sSFR and mass-weighted age}
\label{Results_radial_profiles}

Figure \ref{Radial_profiles} displays the normalized median profiles $\rm \Sigma_{\star}$, $\rm \Sigma_{SFR}$, $A_{\rm v}^{\rm ISM}$, sSFR and mass-weighted age  inferred by our SED fitting procedure. The median $\rm \Sigma_{\star}$ profiles of $\rm RedSFGs$ and QGs are almost identical, leading to similar half-mass radii that are about $37\%$ smaller than those of $\rm BlueSFGs$. The latter show a shallower increase in $\rm \Sigma_{\star}$ toward the center, indicating a more extensive stellar profile. This indicates that the $\rm RedSFG$ population has developed a bulge characteristic of the QG population.

According to the $\rm \Sigma_{SFR}$ distribution of QGs, they no longer form stars at all radii, which is not the case for $\rm BlueSFGs$ showing a steady decrease toward their outskirts. On the other hand, the SFR profile of $\rm RedSFGs$ is similar in the core to that of $\rm BlueSFGs$, but shows a steeper gradient, meaning that these galaxies are still in the process of building up their central stellar bulge.

To compare past star formation, measured by $\rm \Sigma_{\star}$, with current star formation, given by $\rm \Sigma_{SFR}$ as a function of radius, we measured the sSFR profiles of our galaxies. This gives us indication on ongoing evolution of their structure. The constant sSFR at all radii for $\rm BlueSFGs$ and $\rm RedSFGs$ indicates that both populations will not be changing their structures much in the near future. On the other hand, QGs have suppressed star formation in their cores that may indicate a stabilization of their central regions while the remaining gas is depleted in their outskirts. Since the $\rm \Sigma_{SFR}$ profile of the QGs seems to be constant, their sSFR profile reflects the $\rm \Sigma_{\star}$ profile. The mass-weighted age profiles of $\rm RedSFGs$ are decaying in their cores similarly to $\rm BlueSFGs$ while being offset to older stellar populations.  Nevertheless, this $\rm RedSFGs$ age distribution is not at the QGs level. This indicate that $\rm RedSFGs$ constitute a population of SFGs that have lived long enough to develop an older, and more massive bulge than more typical $\rm BlueSFGs$.

We see an $A_{\rm v}^{\rm ISM}$ gradient toward the center of the $\rm BlueSFGs$ and $\rm RedSFGs$ indicating an increasing concentration of dust. However, while this gradient is shallow for $\rm BlueSFGs$, it is really steep for $\rm RedSFGs$. On the contrary, QGs do not exhibit any $A_{\rm v}^{\rm ISM}$ radial gradient, and at all radii have $A_{\rm v}^{\rm ISM}$ value consistent with very little dust/gas content. The lack of star formation in these QGs is thus mostly due to the lack of gas content for their mass, rather than a low star formation efficiency. These observations of the $A_{\rm v}^{\rm ISM}$ profile definitively highlight the particularity of $\rm RedSFGs$ as outliers of the star-forming galaxy population with their high attenuation characterized by their high dust concentration.

To study this apparent bimodality between $\rm RedSFGs$ and $\rm BlueSFGs$ as a function of stellar mass, we characterize the slope of the radial profiles presented in Fig.~\ref{Radial_profiles} but grouping galaxies in several stellar mass bins. The slopes of these radial profiles are in this analysis simply approximated by a linear fit of the three innermost annuli, and we named $C_{\rm Mass}$, $C_{\rm SFR}$, $C_{A_{\rm v}^{\rm ISM}}$, $C_{\rm sSFR}$ and $C_{\rm Age}$ the measured slopes that we interpret as the concentrations of these physical quantities. Figure \ref{Slopes} shows that the concentration of stellar mass increases with galaxy mass for $\rm BlueSFGs$ and $\rm RedSFGs$ while QGs are found to have a constant concentration, although probing a relatively narrow mass range. However, we note that the most massive QG exhibit a slightly lower stellar mass concentration which could be interpreted as a sign of recent mergers, possibly dry mergers, which would not increase the stellar mass concentration in the center as in situ star formation does, but rather in their outskirts \citep{naabPropertiesEarlyTypeDry2006, nipotiDRYMERGERSFORMATION2009}. The projected surface density of stars in $\rm RedSFGs$ with ${\rm log}(M_{\ast \rm median}/M_\odot)> 10.5$ exhibits the same concentration as QGs, both being higher than the one of $\rm BlueSFGs$. This suggests that if massive $\rm RedSFGs$ stop forming stars they will end up with the same spatial configuration as QGs. Therefore, from a morphological perspective, $\rm RedSFGs$ could be the progenitors of QGs. In contrast, the most massive $\rm BlueSFGs$ do not show such a concentration of mass, showing that without invoking a violent, major merger event, they feasibly cannot  be the progenitors of QGs. However, the dispersion of high-mass $\rm BlueSFGs$ shows that some SFGs (two in this case) have mass profiles similar to those of $\rm RedSFGs$ and QGs while lacking the $A_{\rm v}^{\rm ISM}$ concentration characteristic of $\rm RedSFGs$. This could indicate an alternative path to passivity. Nevertheless, the bimodality between $\rm BlueSFGs$ and $\rm RedSFGs$ is very significant when looking at the $A_{\rm v}^{\rm ISM}$ concentration, with $\rm RedSFGs$ having more concentrated $A_{\rm v}^{\rm ISM}$ at all masses. This result demonstrates that the $\rm BlueSFGs$/$\rm RedSFGs$ bimodality is not only linked to stellar mass, but that there is a second-order driver of this bimodality, namely the distribution of dust in galaxies at these redshifts that indicates a compaction event building the bulge in situ.

We also observe a drop of the SFR, $A_{\rm v}^{\rm ISM}$, and sSFR concentrations for the most massive $\rm RedSFGs$, to the extent that their properties reassemble those of QGs. Coupled with the increased mass-weighted age concentration indicating an older stellar population in the core, this suggests that for these massive $\rm RedSFGs$ with highly concentrated stellar component, an inside-out quenching mechanism is at play.

\section{Discussion}
\label{Discussion}

We find that above ${\rm log}(M_\ast/M_\odot)> 10$, where all sub-populations are complete, $\rm RedSFGs$ exhibit steeper radial gradients in stellar mass ($\rm \Sigma_{\star}$) and star formation ($\rm \Sigma_{SFR}$) than their parent population of $\rm BlueSFGs$. In addition, we observe a similar stellar concentration between $\rm RedSFGs$ and QGs, particularly for $\rm RedSFGs$ with ${\rm log}(M_\ast/M_\odot)> 10.5$. We would add, however, that the $\rm \Sigma_{\star}$ and $\rm \Sigma_{SFR}$ observed from $\rm RedSFGs$  are located at the upper limit of the $1\sigma$ dispersion of $\rm BlueSFGs$ and that these results could therefore become more significant if the number statistic were higher. These trends therefore indicate that this population has and is still assembling a massive bulge in its core, which is not the case for even the most massive $\rm BlueSFGs$. This bulge growth is probably closely related to the $A_{\rm v}^{\rm ISM}$ gradient, which is steeper for $\rm RedSFGs$ at all masses compared to $\rm BlueSFGs$. If we relate the dust obscuration to the gas content of galaxies, our results suggest that $\rm RedSFGs$ have undergone a phase of compaction of their gas content, which has been channeled into their cores, producing a high $A_{\rm v}^{\rm ISM}$ and SFR gradients. This mechanism therefore allows for compact, MS-like star formation in their cores, particularly around ${\rm log}(M_\ast/M_\odot)\sim10. 5$, where the concentration of $A_{\rm v}^{\rm ISM}$ is highest, which would lead to efficient consumption of the gas into stars and consequently to quenching. We further note that the SFR concentration of $\rm RedSFGs$ derived in this study is likely a lower limit of the true $\rm \Sigma_{SFR}$ of this population. Indeed, while CF00 seems to better recover the true SFR of massive and dusty galaxies (see Sect.~\ref{Stellar mass, star formation rate and attenuation law}), the lack of resolved FIR data does not allow for an accurate determination to be made for the SFR distribution \citep{daddiMultiwavelengthStudyMassive2007,reddyGOODSHerschelMeasurementsDust2012}. This could have a particularly high impact at the center of $\rm RedSFGs$, which is heavily shrouded by dust. The SFR concentration could then be even more pronounced than that observed in this work. 

\begin{figure}
\includegraphics{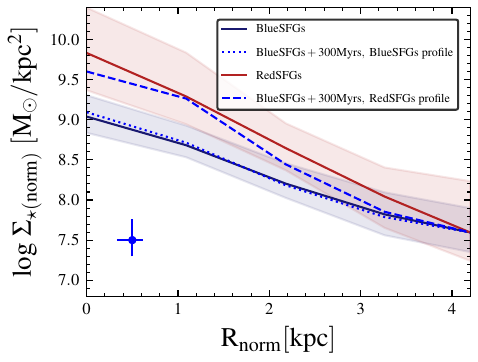}
\caption{Median radial profiles of $\rm \Sigma_{\star}$ for $\rm BlueSFGs$ (in solid blue line) and $\rm RedSFGs$ (in solid red line) with their associated uncertainties represented as blue and shaded regions, respectively. The blue dotted curve represent the radial evolution of $\rm \Sigma_{\star}$ for $\rm BlueSFGs$ over the next 300 Myrs, derived by applying their median radial profile of star formation surface density ($\rm \Sigma_{SFR}$), assumed constant during this period. Additionally, the evolution of the median $\rm \Sigma_{\star}$ radial profile of $\rm BlueSFGs$ over the next 300 Myrs considering the $\rm \Sigma_{SFR}$ profile of $\rm RedSFGs$ is displayed as a dashed blue curve. Each profile is normalised to the $\rm \Sigma_{\star}$ at 4 kpc of $\rm RedSFGs$. We display the typical median uncertainty of the dotted and dashed profiles by a blue error bar. }  
\label{Future_SFGs.pdf}
\end{figure}

The question of whether or not a compaction phase is necessary to explain the concentration of stellar mass observed in $\rm RedSFGs$ is addressed in Fig.~\ref{Future_SFGs.pdf}. Here, we calculate the evolution of the radial distribution of stars from $\rm \Sigma_{SFR}$ in each annulus and show that $\rm BlueSFGs$ will not naturally transform into $\rm RedSFGs$ if they maintain their radial $\rm \Sigma_{SFR}$ profile over the next 300 Myrs (blue dotted curve). Instead, a significant compaction event with a steeper $\rm \Sigma_{SFR}$ profile is required for $\rm BlueSFGs$ to evolve into $\rm RedSFGs$ (blue dashed curve). The duration of this compaction phase, around 300 Myrs, corresponds to the mass-weighted median age profile observed at the center of the $\rm RedSFGs$ and QGs and is consistent with the 500 Myrs quenching timescale in compact SFGs at $z\sim2$ found in  \citet{vandokkumFormingCompactMassive2015}. This suggests that $\rm RedSFGs$ are the result of the evolution of $\rm BlueSFGs$ which experienced a compaction event over a duration of around 300 Myrs.

Figure \ref{Slopes} (lower left panel) also shows that while most $\rm BlueSFGs$ and $\rm RedSFGs$ do not exhibit any strong radial sSFR gradient (indicating that the stellar structure of these populations will not be modified much by current star formation activity), the most massive $\rm RedSFGs$ break this trend by displaying a similar sSFR profile to the QGs. We interpret this population as being depleted of gas at its center (decrease in its $A_{\rm v}^{\rm ISM}$ concentration), while continuing to form stars at its periphery. This leads to a decrease in its stellar mass concentration and a converging to that of the QG population.

Finally, the median mass of the $\rm RedSFG$ population (${\rm log}(M_\ast/M_\odot)\sim 10.5$), together with its dominance above this mass, and our observation that the most massive $\rm RedSFGs$ exhibit a typical QG stellar mass profile, point in the direction of a major compaction event that allows for the onset of the in situ growth of a bulge. In this scenario, some high redshift $\rm BlueSFGs$ are experiencing this compaction event after spending some time on the MS and reaching a typical stellar mass of ${\rm log}(M_\ast/M_\odot)\sim 10.5$. This major compaction is assumed to be “wet” according to \citet{dekelWetDiscContraction2014, zolotovCompactionQuenchingHighz2015a} so that during this phase, angular momentum loss allows cold gas to be efficiently channeled into galaxy cores, producing MS-type star formation with optically thick dust and ISM enrichment. What triggers such a compaction phase is beyond the scope of this article, but scenarios such as a major merger or consecutive minor mergers \citep{dekelDissipativeMergerProgenitors2006,covingtonRoleDissipationScaling2011}, counter-rotating cold gas accretion \citep{danovichFourPhasesAngularmomentum2015}, recycling fountains \citep{chabanierFormationCompactGalaxies2020} or satellite tidal compression \citep{renaudStarburstsTriggeredIntergalactic2014} are advocated by simulations and facilitate efficient in situ bulge mass growth. The result is a compact mass profile consistent with that of passive galaxies, once all the gas in the galaxies has been depleted and star formation has stopped. To reconcile the high dust content of $\rm RedSFGs$ with the small but still existing amount of dust in $z \sim 2$ QGs, it is worth noting that the typical time scale for dust removal in post-starburst galaxies is $\sim 2$ Gyrs \citep{michalowskiFateInterstellarMedium2019}. Therefore, the most massive $\rm RedSFGs$ could be in this post-starburst phase, removing their dust while keeping a highly concentrated mass profile, then converging toward the properties of QGs at $z \sim 2$.

The number density of $\rm RedSFGs$ ($n\sim 1.23 \times 10^{-4} \rm Mpc^{-3}$) is of the same order of magnitude as the one of QGs at $z \sim 2$ ($n\sim 10^{-4} \rm Mpc^{-3}$) \citep{mcleodEvolutionGalaxyStellarmass2021}. This suggests that $\rm RedSFGs$ at $3<z<4$ are good candidate for being the progenitor of typical massive QGs at $z\sim 2$ \citep{mcleodEvolutionGalaxyStellarmass2021} and is also consistent with the fact that the bulk of QGs at $z\sim 2$ are compact \citep{vanderwel3DHSTCANDELSEvolution2014,vandokkumFormingCompactMassive2015}.

We also note that the typical stellar mass (${\rm log}(M_\ast/M_\odot)\sim 10.5$) of $\rm RedSFGs$ corresponds to the critical mass defining the bimodality between $\rm BlueSFGs$ and red QGs in the Local Universe \citep{baldryQuantifyingBimodalColorMagnitude2004}. This so-called critical mass was shown to be explained by a wet compaction scenario \citep{zolotovCompactionQuenchingHighz2015a,tacchellaEvolutionDensityProfiles2016, dekelOriginGoldenMass2019}. In this scenario, the wet compaction phase occurs close to this stellar mass, as it is confined by two main physical mechanisms: supernova feedback and virial shock heating of the circumgalactic medium (CGM). On the one hand, for $M_\ast < M_{\rm crit}$, the galaxies' potential well is not deep enough to retain the gas expelled by supernova feedback or to allow for a significant portion of the gas to heat up, making them particularly effective at suppressing star formation and consequently stopping the gas compaction. On the other hand, for $M_\ast > M_{\rm crit}$, the CGM surrounding the halos is heated by viral shocks that warm the incoming cold gas coming from the intergalactic medium via filaments. The result is a small amount of cold gas, which is not conducive to efficient star formation. It is also suggested that the growth of a super-massive black hole in such a massive system is possible and would guarantee long-term quenching through AGN feedback \citep{weinbergerSupermassiveBlackHoles2018} once star formation has stopped. According to \citet{perez-gonzalezCEERSKeyPaper2023} and \citet{barrufetQuiescentDustyUnveiling2024} it is possible that ${\sim} 10\%$ of $\rm RedSFGs$ contain a highly obscured AGN.

Overall, this scenario seems consistent with our observations of this red population and suggests that $\rm RedSFGs$ can be defined as a phase in galaxy evolution that creates a bimodality between blue and red SFGs. This bimodality in SFGs in place at $z=3-4$ defines a primeval bimodality that  leads to the formation of today's massive ellipticals.  

\section{Summary}
\label{Summary}

In this work, we study the transition from SFGs to QGs by investigating the morphology of a color-selected, massive, and dust-obscured population of galaxies, faint in the observed UV/optical, but bright in the NIR -- making them particularly red (hence, the nomer $\rm RedSFGs$ used in this work). We focused on this population in the context of a mass-complete sample of 188 galaxies at $3<z<4$ with ${\rm log}(M_\ast/M_\odot)>9.6$, cleaned from potential AGN contaminants, point sources, and galaxies with uncertain photometric redshift determination in the JWST/CEERS field.

Among our mass-complete sample, ${{\sim}} 17\%$ of the galaxies are classified as red $\rm RedSFGs$, ${\sim} 76\%$ are blue $\rm BlueSFGs$, and ${\sim} 7\%$ are classified as QGs. We fit the resolved SED of each galaxy and derived its median $\rm \Sigma_{\star}$, $\rm \Sigma_{SFR}$, sSFR, $A_{\rm v}^{\rm ISM}$, and mass-weighted age profiles. Our conclusions can be summarized as follows:
\begin{enumerate}
\item
$\rm RedSFGs$ and QGs taken together account for over $50\%$ of galaxies at ${\rm log}(M_\ast/M_\odot)>10.5$ and this fraction increases with mass. They make up $> 80\%$ at ${\rm log}(M_\ast/M_\odot)\sim 11$ with a large predominance of $\rm RedSFGs$ for ${\rm log}(M_\ast/M_\odot)>10.6$, where they represent $50\%$ of all galaxies.
\item
The radial profiles confirm the compactness of the $\rm RedSFGs$ stellar mass distribution, which is similar to that of the QGs. This suggests that this population is the link in terms of morphological transition between disk-dominated $\rm BlueSFGs$ and bulge-dominated QGs. Their mass-weighted age profiles are also shifted to older ages compared to $\rm BlueSFGs$, while showing an overall trend similar to them. It confirms that $\rm RedSFGs$ are SFGs on their way to becoming QGs after  a certain amount of time on the MS.
\item
The $\rm \Sigma_{SFR}$ and dust attenuation profiles of $\rm RedSFGs$ also show a high concentration and indicate a gas compaction event. These profiles suggest that we are witnessing the massive bulge growth necessary to explain the compact stellar morphology of $\rm RedSFGs$. In addition, we tested whether $\rm BlueSFGs$ were able to obtain the $\rm \Sigma_{\star}$ profiles of $\rm RedSFGs$ in less than 300 Myrs by applying the $\rm \Sigma_{SFR}$ profile of $\rm RedSFGs$ to them. We have found that only the application of $\rm RedSFGs$ $\rm \Sigma_{SFR}$ profile can produce a $\rm \Sigma_{\star}$ distribution similar to that of $\rm RedSFGs$. This reinforces the need for a compaction event happening in situ to become $\rm RedSFGs$ and, subsequently, QGs. This compaction event  then goes on to create a morphological bimodality between $\rm BlueSFGs$ and $\rm RedSFGs$.
\item
We have investigated the evolution with mass of this bimodality and found that the dust attenuation, $\rm \Sigma_{\star}$, and $\rm \Sigma_{SFR}$ are more concentrated at all masses in $\rm RedSFGs$, with a peak of these concentrations for ${\rm log}(M_\ast/M_\odot)\sim10.5$. The fact that the most massive $\rm BlueSFGs$, which do not show any sign of compaction, do not reach the stellar mass concentration of QGs, proves that increasing stellar mass alone cannot drive a SFG toward quiescence and that compaction is required.
\item
The most massive $\rm RedSFGs$ in our sample, although showing a highly concentrated bulge, strong dust attenuation, and MS-type star formation, also show a significant decrease in $\rm \Sigma_{SFR}$, $A_{\rm v}^{\rm ISM}$, and sSFR concentrations corresponding to suppressed star formation in their cores which contrasts with lower mass $\rm RedSFGs$. We interpret these deviations from the general mass trends as a sign that these galaxies are engaged in an inside-out quenching process.
\item
Our results suggest that the $\rm RedSFG$ population is linked to the critical mass ${\rm log}(M_{\ast \rm crit}/M_\odot)\sim 10.5$ defining the bimodality between QGs and SFGs in the Local Universe. This validates the nature of this population as the primeval link between SFGs and QGs.
\end{enumerate}
  
Our findings can be related to the galaxy evolution scenario suggested in \citet{behrooziAverageStarFormation2013}, \citet{dekelOriginGoldenMass2019} by assuming that $\rm RedSFGs$ define a population of SFGs that undergo a compaction phase. They experience the final compaction of their gas content once they reach the critical mass ${\rm \log}(M_{\ast {\rm crit}}/M_{\odot})\sim 10.5$, leading to inside-out quenching. This means that the so-called bimodality between local SFGs and QGs originates in a bimodality already existing at $z=3-4$ between SFGs that have experienced ($\rm RedSFGs$) or are still experiencing a compaction event and those that have not ($\rm BlueSFGs$). We suggest that this primeval bimodality takes its seed in relatively rapid compaction events (<300 Myrs) that lead to a short depletion timescale to transform the gas into stars in a concentrated, dust-shrouded region. The dust channeled into the core and produced by star formation activity is trapped in the galaxy's gravitational potential, producing a steep $A_{\rm v}^{\rm ISM}$ gradient characteristic of the $\rm RedSFG$ phase. To put  better constrains on the dust and gas content in this obscured population, deep and high-resolution imaging with ALMA would be necessary. Combined with multiwavelength photometry, especially with MIRI, this would allow better stellar mass estimates of these obscured system and shed light on the possible AGN contribution in their cores. Spatially resolved spectroscopy could also be of interest to study the metallicity and age gradients as well as the resolved SFH and velocity profiles of these galaxies. In particular, studies of the kinematic of $\rm RedSFG$ would allow us to measure their three-dimensional stellar distribution and explore their triaxial properties. This three-dimensional information is vital for testing and confirming the in situ bulge formation scenario that we are highlighting in this work. Putting all these informations together, we will be able to lift the veil on the physical process underlying the major compaction phase in the early Universe.

\begin{acknowledgements}
MT acknowledges support from CNES. This work was supported by the Programme National Cosmology et Galaxies (PNCG) of CNRS/INSU with INP and IN2P3, co-funded by CEA and CNES.
MF acknowledges financial support from the European Union’s Horizon 2020 research and innovation programme under the Marie Sklodowska-Curie grant agreement No 101148925. MT acknowledges the following open sources softwares used in this work : \texttt{Numpy} \citep{harrisArrayProgrammingNumPy2020}, \texttt{Astropy} \citep{astropycollaborationAstropyCommunityPython2013,astropycollaborationAstropyProjectBuilding2018,astropycollaborationAstropyProjectSustaining2022}, \texttt{Photutils} \citep{larry_bradley_2024_12585239} and \texttt{SciencePlots} \citep{johngarrettGarrettj403SciencePlots2112023}.

\end{acknowledgements}

\bibliographystyle{aa}
\bibliography{main}

\end{document}